\documentclass{article}

\PassOptionsToPackage{numbers, compress}{natbib}

     \usepackage[preprint]{neurips_2024}

\usepackage[utf8]{inputenc} 
\usepackage[T1]{fontenc}    
\usepackage[breaklinks=true]{hyperref}       
\usepackage{url}            
\usepackage{booktabs}       
\usepackage{amsfonts}       
\usepackage{nicefrac}       
\usepackage{microtype}      
\usepackage{xcolor}         

\usepackage{amsmath}        
\usepackage{graphicx}

\usepackage{algorithm2e}

\usepackage{pdflscape}
\usepackage{afterpage}

\newcommand*{\textcite}{\citet}

\usepackage{comment}

\usepackage{soul}

\usepackage{etoolbox}
\robustify{\st}
\robustify{\hl}
\sethlcolor{white}
\setstcolor{red}

\usepackage[sectionbib]{bibunits}
\defaultbibliographystyle{plainnat}
\defaultbibliography{Library.bib}

\soulregister\textcite7
\soulregister\ref7
\soulregister\cite7   

\title{Data-Error Scaling\hl{ Laws} in Machine Learning on Combinatorial Mutation-prone Sets: \hl{Proteins }and Small Molecules}

\author{%
Vanni Doffini$^{1,2,3,}$\thanks{\texttt{vanni.doffini@unibas.ch}} \quad O. Anatole Von Lilienfeld$^{4,5,6}$  \quad Michael A. Nash$^{1,2,3,}$\thanks{\texttt{michael.nash@unibas.ch}}\\
$^1$\textit{University of Basel} \quad
$^2$\textit{ETH Zurich} \quad
$^3$\textit{Swiss Nanoscience Institute} \\
$^4$\textit{University of Toronto}\quad
$^5$\textit{Vector Institute}\quad
$^6$\textit{TU Berlin}
}

\begin{document}

\maketitle

\begin{abstract}
        We investigate trends in the data-error scaling \hl{laws} of machine learning (ML) models trained on discrete combinatorial spaces that are prone-to-mutation, such as proteins or organic small molecules. We trained and evaluated kernel ridge regression machines using variable amounts of \hl{computational and experimental} training data. Our synthetic datasets comprise\hl{d} i) two naïve functions based on many-body theory; ii) binding energy estimates between a protein and a mutagenised peptide; and iii) solvation energies of two 6-heavy atom structural graphs\hl{, while the experimental dataset consisted of a full deep mutational scan of the binding protein GB1}. In contrast to typical data-error scaling\hl{ laws}, our results showed discontinuous monotonic phase transitions during learning, observed as rapid drops in the test error at particular thresholds of training data. We observed two learning regimes, which we call saturated and asymptotic decay, and found that they are conditioned by the level of complexity (i.e. number of mutations) enclosed in the training set. We show that during training on this class of problems, the predictions were clustered by the ML models employed in the calibration plots. Furthermore, we present  an alternative strategy to normalize learning curves (LCs) and \hl{introduce }the concept of mutant\hl{-}based shuffling. This work has implications for machine learning on mutagenisable discrete spaces such as chemical properties or protein phenotype prediction, and improves basic understanding of concepts in statistical learning theory. 
\end{abstract}

\begin{bibunit}

\section{Introduction}

Machine learning (ML) has garnered significant interest in recent years, particularly with the release of generative models such as BERT \cite{VD_LLM_BERT}, (Chat-)GPT \cite{VD_Extra_GPT4}, Gemini \cite{VD_LLM_Gemini}, LlaMA \cite{VD_LLM_LlaMa,VD_LLM_LlaMa2}, Mistral \cite{VD_LLM_Mistral7B}, DALL-E \cite{VD_T2I_DALLE,VD_T2I_DALLE2,VD_T2I_DALLE3}, and Stable Diffusion \cite{VD_T2I_StableDiffusion, VD_T2I_StableDiffusion2} to the general public. Similarly, the introduction of Alphafold \cite{VD_ML4ProtSpace_41,MN_MachineLearningProteinStucture_01,VD_Attention_02} represented a turning point in protein science \cite{VD_ML4ProtSpace_39,VD_ML4ProtSpace_40}, providing researchers with the ability to predict 3D structures \cite{VD_ML4ProtSpace_01,VD_ML4ProtSpace_02} of single proteins or even protein complexes from primary sequences \cite{VD_ML4ProtSpace_07}. Nevertheless, predicting the effects of mutations on protein phenotype \cite{VD_ML4ProtSpace_24} is a monumentally challenging task \cite{VD_AlphaFoldLimits_01, VD_AlphaFoldLimits_02} due to the complexity of the problem (epistasis) \cite{VD_ML4ProtSpace_18,VD_ML4ProtSpace_15}, the wide range of protein phenotypes of potential interest, and the dearth and cost of high-quality training data.

\begin{figure}[htb!]
  \centering
  \includegraphics[width=\textwidth]{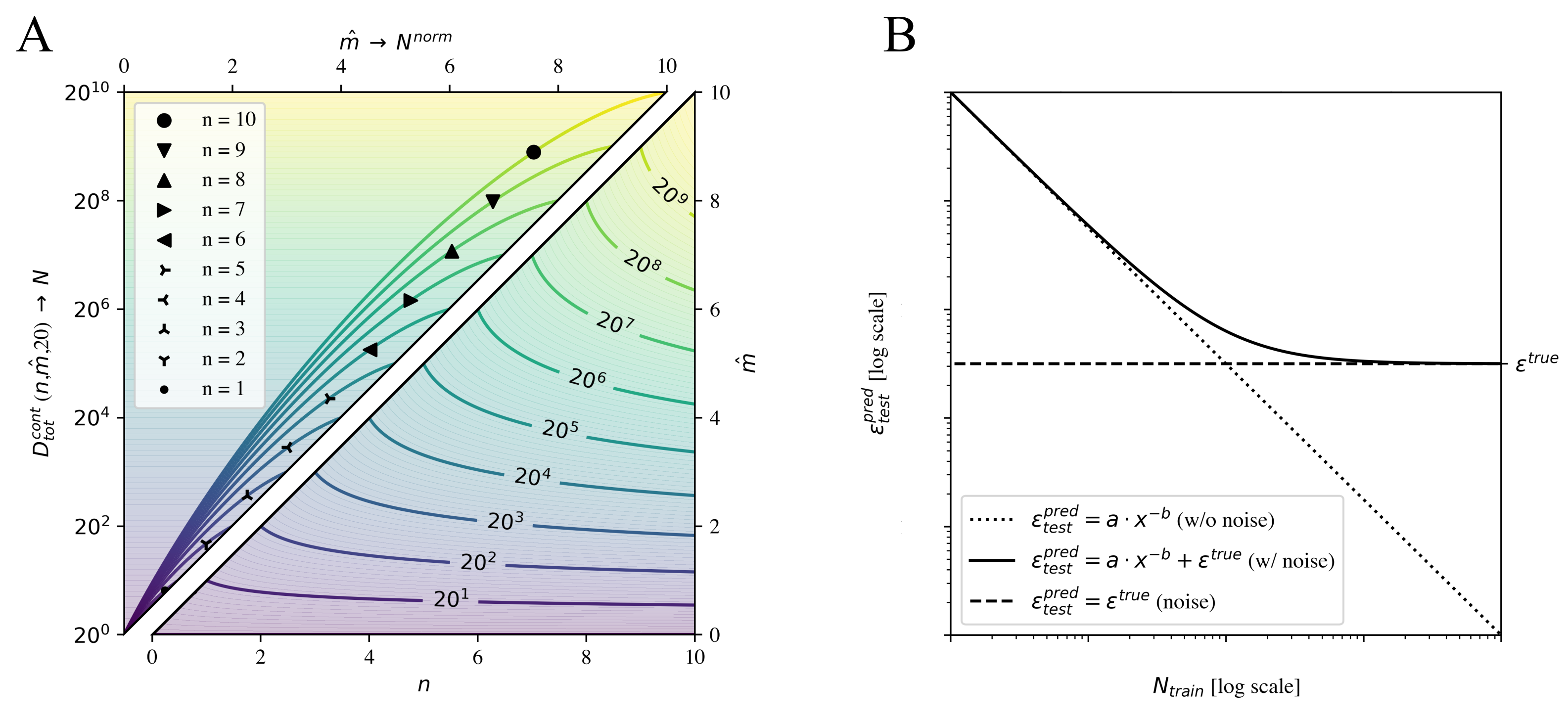}
  \caption[Size of cumulative combinatorial space for a linear graph and ideal learning curve.]{Size of cumulative combinatorial space for a linear graph (e.g., a peptide)\hl{ and ideal LC}. \hl{(A)} The number of cumulative combinations ($D_{tot}^{cont}$) is shown while changing the length of the linear graph ($n$, number of amino acids in chain) and the number of mutations in the sequence ($\hat{m}$). The vocabulary size was kept constant (20). \hl{(B) Example of an idealized LC (i.e., following a single power-law) for a classic ML problem with non-mutation-prone inputs. The effect of homoscedastic noise (dashed line) on the ideal behavior is shown: dotted line = no noise; solid line = with noise. Numerical values on the axes are intentionally omitted to highlight the illustrative and idealized nature of this subfigure.}}
  \label{fig:combinatorial_space_scale}
\end{figure}

One meaningful difference in this context exists between global (e.g., de-novo protein design, Alphafold structure prediction) and local (e.g., variant libraries, directed evolution) optimization \cite{VD_ML4ProtSpace_34}. For global optimization, a large number of diverse proteins need to be cataloged, annotated or labeled. In contrast, for local optimization a combinatorically large number of mutant sequences (Fig. \ref{fig:combinatorial_space_scale}\hl{A}) with high similarity to a single parent sequence (wild-type, WT) must be screened and assigned phenotype values. Even with state-of-the-art experimental techniques such as high-throughput screening \cite{VD_ML4ProtSpace_25,MN_Next-GenSequencingForEnzymeEvolution_04}, next generation sequencing (NGS) \cite{MN_Next-GenSequencingForEnzymeEvolution_10,MN_Next-GenSequencingForEnzymeEvolution_06} and deep-mutational scanning (DMS) \cite{VD_MyArticles_01,VD_MyArticles_02,MN_DMS_01}, or automated laboratories \cite{VD_ML4ProtSpace_33,VD_ML4ProtSpace_26},  the fraction of variants that can be analyzed is miniscule compared to the size of potentially interesting sequence space. 

De-novo protein design is broader but coarser, while mutant-based design is task-specific and less generalizable. Nevertheless, scientists have successfully implemented ML for both tasks, generating new proteins \cite{VD_ML4ProtSpace_16,VD_ML4ProtSpace_17} and enhancing the biological fitness of variants derived from parent sequences \cite{MN_MachineLearningDirectedEvolution_09,VD_ML4ProtSpace_11,VD_ML4ProtSpace_18,VD_ML4ProtSpace_29}. The latter has been especially useful for enhancing directed evolution \cite{VD_ML4ProtSpace_12,VD_ML4ProtSpace_37,VD_ML4ProtSpace_35,VD_ML4ProtSpace_36}, an experimental approach where specific traits are enhanced through iterative screening/selection and enrichment of variant sequences possessing beneficial mutations \cite{VD_DirectedEvolution_01,VD_DirectedEvolution_02}.

To validate ML-based approaches in protein and molecular engineering, it is important to evaluate how quickly a given model is able to learn a given task. This allows comparisons of performance and efficiency to be made between various models. The learning \hl{dynamics} depends on several parameters, such as the architecture (complexity) of the model, how data are encoded, if feature engineering is employed as well as how the cost of model training scales with input data size. 

One important tool to study, compare and extrapolate the learning process of ML models is the so-called learning curve (LC) \cite{VD_LearningCurves_02,VD_LearningCurves_03,VD_LearningCurves_04}, which is a subset of the neural scaling law concept \cite{VD_NeuralScalingLaw_09,VD_NeuralScalingLaw_08,VD_NeuralScalingLaw_05_preprint,VD_NeuralScalingLaw_02_preprint,VD_NeuralScalingLaw_01}. LCs were introduced in the statistical mechanics field \cite{Book_ML_13,VD_LearningCurves_05,VD_LearningCurves_12,VD_LearningCurves_13} and recently popularized in computational chemistry and physics to assess predictions of quantum machine learning (QML) tasks \cite{VD_ML4ChemSpace_13,VD_ML4Chemspace_11}. The main concept is to train a model consisting of a particular architecture, data structuring, or algorithm by feeding it increasing fractions of the total available training data. For each training instance, a performance metric on an independent and constant (test) dataset is evaluated. Scaling in LCs can usually be modelled as a power law with an offset \hl{(Fig. \ref{fig:combinatorial_space_scale}B)} \cite{VD_LearningCurves_07,VD_LearningCurves_01}, according to

\begin{equation}\label{eq:standard_lc}
\epsilon_{test}^{pred}=a\left(N_{train}\right)^{-b}+c
\end{equation}

where the offset (c) is related to the level of noise in the data \hl{($\epsilon^{true}$)}, \hl{$\epsilon_{test}^{pred}$}represents the model error, and the parameters (a) and (b) describe the learning \hl{dynamics} of the model\hl{, namely where it starts learning (a) and how quickly it learns (b)}. Fitting different architectures and techniques result in different LCs, which can be compared to assess diverse ML models and decide which model presents better scalability. LCs can also be used to quantify if and how much additional data are necessary to reach a certain accuracy. Despite a strong desire in the community to understand on a deeper level how the ML process works, LCs are sparsely used outside theoretical fields. 

In the case of protein phenotype prediction, the learning tasks are of a special kind. Protein sequence space is discrete, with a fixed set of typically 20 amino acids included in the mutational vocabulary of proteins. To the best of our knowledge, there has not been an in-depth study to understand how the combinatorial discreteness of protein sequence space could influence the shape and behavior of learning curves. A similar paradigm can also be applied to ML of small molecule properties where molecules can be ‘mutagenized’ by altering substituent groups, however, the mutational space can be further complicated by changes in topology of the backbone upon mutation. 

To bridge this knowledge gap, we present a\hl{n} empirical study on the scaling behavior of LCs obtained during ML training on different restricted (bio-)chemical input spaces with deterministic fitness functions\hl{ and experimentally measured values}. Using our workflow (Fig. \ref{fig:workflow_overview}), we show that the expected LC trends cannot be described by single power laws, but exhibit periods of accelerated and decelerated learning, a behavior closely related to discontinuous learning \cite{Book_ML_13,VD_LearningCurves_07,VD_NeuralScalingLaw_08,VD_LearningCurves_13}. Moreover, we show that the accelerations observed in the LCs emerge when training examples containing higher numbers of mutations are introduced or upon saturation and exhaustion of training examples containing a certain number of mutations in the training set. These findings could highly impact the way experiments and/or simulations are planned, how information-rich mutational data are most efficiently generated, and how they can be scaled. This is particularly relevant in biology, but can also be extended to other fields where discrete combinatorial prone-to-mutation input spaces are relevant.

\section{Results and Discussion}
\subsection{Workflow Overview} 

\begin{figure}[htb!]
  \centering
  \includegraphics[width=\textwidth]{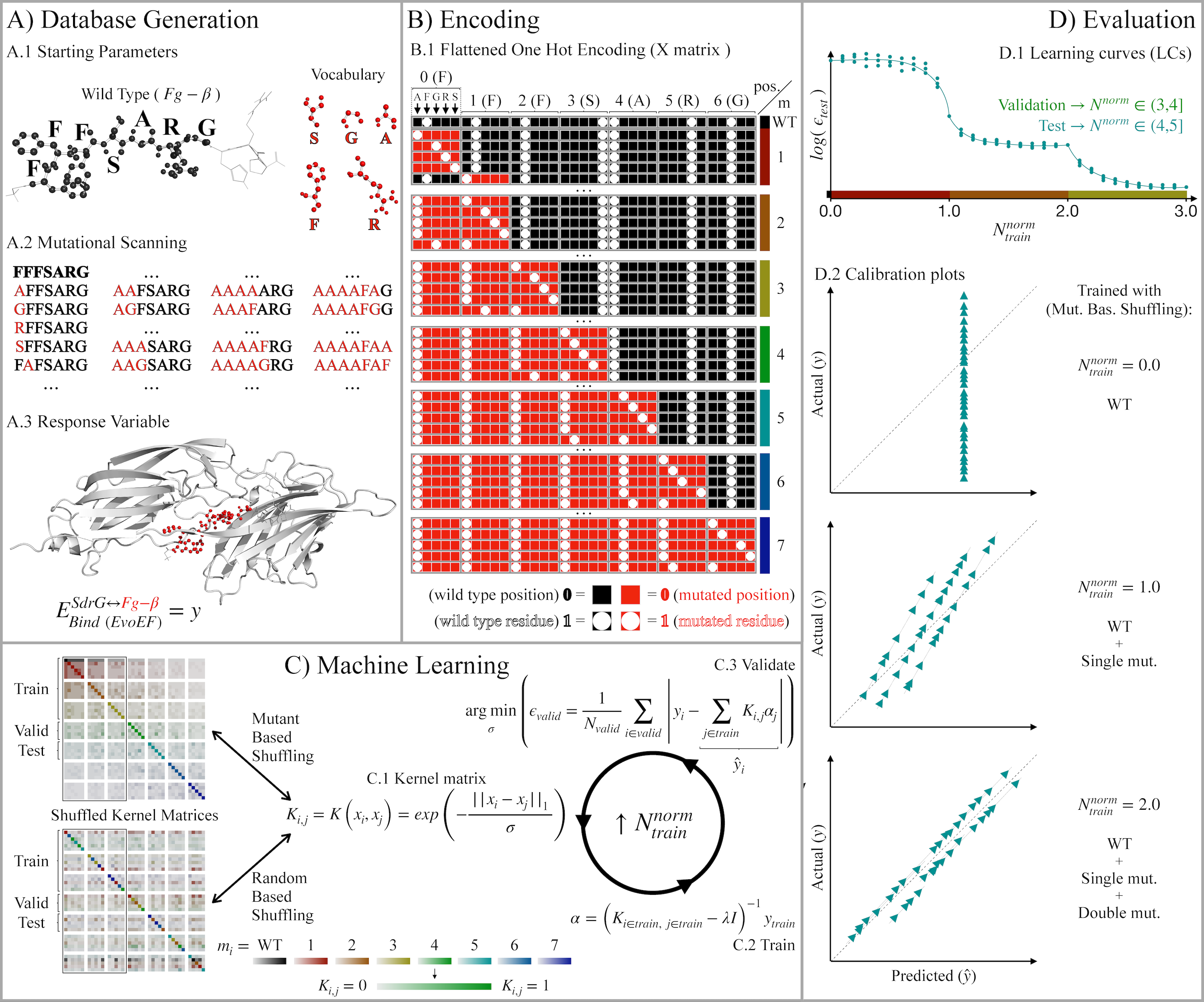}
  \caption[Workflow overview.]{Workflow overview. (A) Database Generation: a table containing all possible mutagenized peptide variants was generated from a starting construct (WT) and a mutational vocabulary. The response variable (binding energy) was computed for each entry. (B) Encoding: the database was converted into a matrix containing numerical values using binary flattened one hot encoding. (C) Machine learning: Laplacian kernel machines were trained using different quantities of data and different shuffling strategies. (D) Evaluation: LCs and calibration plots were used to study the learning process. The amount of information used during training in the scatter plots is reported on the figure.}
  \label{fig:workflow_overview}
\end{figure}

We started by generating two databases containing all possible point mutations on three systems, consisting of a peptide and two small molecules. The systems were (1) Fg-$\beta$, a peptide derived from a portion of the human fibrinogen beta chain (Fig. \ref{fig:workflow_overview}A); (2) hexane; and (3) cyclohexane. Although the term “mutation” is typically utilized for protein variants derived from a parent or WT sequence, here we extended this concept to organic small molecules by substituting one or more heavy atoms, carbon in our case, with nitrogen or oxygen, fixing the hybridization to Sp3 and filling with hydrogens. In contrast to the case with protein mutants where each point mutation is unique, for simple molecules this is not true due to symmetries. Such symmetric mutations can lead to duplicate entries in our molecule databases, however, for simplicity we left such duplicates in place.

We then used different functions to generate a set of response variables for each database entry. For the peptide variant database, we utilized two many-body based functions \cite{Book_ML_04} and an estimator of the binding affinity (EvoEF) \cite{DL_PeptidesBinding_01,DL_PeptidesBinding_02} between the mutagenized peptide and its natural receptor, a protein called SdrG from S. epidermidis \cite{VD_SdrGFgB_1r17} (Fig. \ref{fig:learning_curves}B). For the small molecules in the chemical database, we calculated the free energy of solvation in water at 298K using LeRuLi \cite{LeRuLi, LeRuLi_solvation_energy_01, LeRuLi_solvation_energy_02}, which utilized the Abraham linear solvation energy relationship (LSER) \cite{ LeRuLi_solvation_energy_05, LeRuLi_solvation_energy_04} and the Platts group additivity method \cite{LeRuLi_solvation_energy_03}.

\hl{In addition to these synthetic benchmarks, we included an experimental dataset from a DMS campaign of the B1 domain of protein G (GB1) \cite{VD_DataBio_08} to validate our in-silico observations (see below).}

We then converted each member in the database to a numerical value using standard one-hot-encoding (OHE, Fig. \ref{fig:workflow_overview}B)\hl{ and additional domain specific encodings (i.e., AAindex, z-scores, Coulomb matrix, ECFP4, and ESM embeddings; see Materials and Methods)}. We used Laplacian kernel machines to learn the generated response variables from the OHE matrix using different numbers of training samples (Fig. \ref{fig:workflow_overview}C). The hyperparameter optimization ($\sigma$, kernel scale) was performed via grid search using as validation set a number of datapoints equal to the number of possible quadruple mutations. To test the performance, we used the mean absolute error (MAE) metric on an independent test set containing a number of datapoints equal to the number of possible quintuple mutations. Different shuffling techniques were applied to different portion of the datasets, which included random or mutant based (Fig. \ref{appendix:fig:1}). The latter consisted of sorting the dataset (or a subset) accordingly to the number of mutations contained in each variant after randomly shuffling it.
We then converted each member in the database to a numerical value using standard one-hot-encoding (OHE, Fig. \ref{fig:workflow_overview}B)\hl{ and additional domain specific encodings (i.e., AAindex, z-scores, Coulomb matrix, ECFP4, and ESM embeddings; see Materials and Methods)}. We used Laplacian kernel machines to learn the generated response variables from the OHE matrix using different numbers of training samples (Fig. \ref{fig:workflow_overview}C). The hyperparameter optimization ($\sigma$, kernel scale) was performed via grid search using as validation set a number of datapoints equal to the number of possible quadruple mutations. To test the performance, we used the mean absolute error (MAE) metric on an independent test set containing a number of datapoints equal to the number of possible quintuple mutations. Different shuffling techniques were applied to different portion of the datasets, which included random or mutant based (Fig. \ref{appendix:fig:1}). The latter consisted of sorting the dataset (or a subset) accordingly to the number of mutations contained in each variant after randomly shuffling it.

To study the impact of including more information in the training dataset, we generated the LCs (Fig. \ref{fig:workflow_overview}D.1) and the actual vs. predicted scatter plots, also known as calibration plots, at different training instances (Fig. \ref{fig:workflow_overview}D.2). In order to account for the non-linearity of the combinatorial nature of our mutagenizable systems (Fig. \ref{fig:combinatorial_space_scale}), we developed an alternative way to scale the number of datapoints in the LCs (see Supporting Information, Materials and Methods). The number of datapoints was converted using a formula related to the beta function. To give an example, any number of single mutants (plus WT) can be mapped to a number between between zero (one entry) and one (a number equal to all possible single mutants plus the WT). Each unit along the x-axis of the learning curve therefore represented a number of training examples that had saturated mutations of a given order (i.e., 1st order mutations, double mutants, etc.).

\subsection{Phase-transitioning learning curves (LCs)}

\begin{figure}[htb!]
  \centering
  \includegraphics[width=\textwidth]{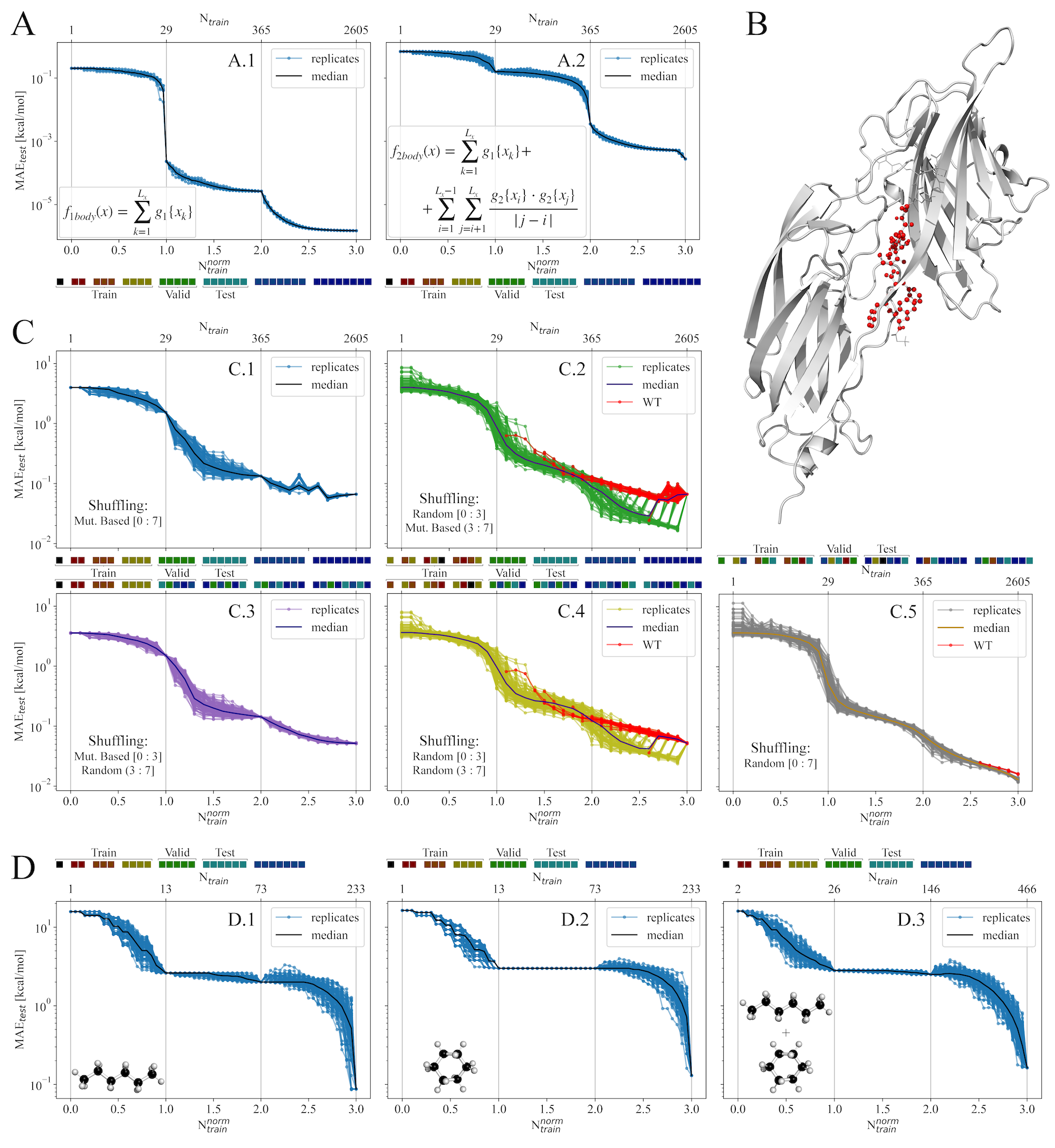}
  \caption[Learning Curvers (LCs) for different discretized spaces and functions.]{LCs for different discretized spaces and functions. (A) 1-body (left) and 2-body (right) naïve functions. (B) Fg-$\beta$ (red) / S. epidermidis adhesin SdrG (greyscale) complex. (C) Binding energy function results using different shuffling. (D) Solvation energies results on different structures (6 heavy atoms linear, cyclic or the combination of both). The shuffling strategies are specified when relevant. If not explicitly specified, mutant-based shuffling is applied to the whole dataset. The presence of WT in the training set is reported in red only if the shuffling strategy did not automatically set it as the first entry.}
  \label{fig:learning_curves}
\end{figure}

The LCs of the biological and chemical systems described above are shown in Fig. \ref{fig:learning_curves}. We started by analyzing the performance of our kernel machines on two simple functions based on many-body theory (Fig. \ref{fig:learning_curves}A). We observed two possible regimes arising from feeding the model with mutant-based shuffled data. We called such behaviors “asymptotic” and “saturated” decay. Asymptotic decay was observed for higher order mutations and consisted of a trend similar to the one typically reported, which followed a power law decay tending asymptotically to an offset. In contrast to typical LCs reported in the literature, this behavior was observed strictly within specific regions where a certain number of mutations were enclosed in the training dataset, but not across the whole LC. In regions not characterized by asymptotic decay, we observed saturated decay. In this scenario, the MAE remained constant in the initial phase and dropped suddenly at the point where most variants with a specific number of mutations had been included (saturated) in the training dataset. In addition, we found that the regions where saturated decay was observed correlated with the order of the function used to generate the response variables. In the case of a linear function (1-body-term) the saturated decay was limited to the single mutant region (Fig. \ref{fig:learning_curves}A.1, $N_{train}^{norm} \in (0,1]$), while with a nonlinear function (second order) two saturation decays were observed (Fig. \ref{fig:learning_curves}A.2, $N_{train}^{norm} \in (0,1]$ and $N_{train}^{norm} \in (1,2]$).

To determine if this scaling behavior generalized to other functions, we used EvoEF to estimate the affinity (binding energy) between a mutagenized peptide (Fg-$\beta$) and a known target protein (SdrG, Fig. \ref{fig:learning_curves}B). We investigated the influence of the mutant shuffling strategies applied to the dataset on the LCs (Fig. \ref{fig:learning_curves}C). The training data were shuffled using a mutation-based scheme and the training set was filled in a hierarchical way with up to triple mutants (Fig. \ref{fig:learning_curves}C.1). The results showed similar behavior to what was previously observed in the 1-body term case (Fig. \ref{fig:learning_curves}A.1). We observed saturated decay in the single mutant region of the LC ($N_{train}^{norm} \in (0,1]$), and two asymptotical decays in the double ($N_{train}^{norm} \in (1,2]$) and triple mutants regions ($N_{train}^{norm} \in (2,3]$). The latter showed the presence of noise with some spikes in the MAE. This was mitigated by randomly shuffling the variants containing more than 4 mutations, forcing the validation and the test to include all possible mutations above triples (included only in the training set) and making both similarly distributed (Fig. \ref{fig:learning_curves}C.3). The shuffling strategy of the data which constituted the training set seemed to have the highest impact on the LCs. In fact, if the hierarchical order of the mutations was not maintained in the training set, a new behavior arose. This result was independent from the methodologies employed, irrespective of whether the dataset was randomly shuffled (Fig. \ref{fig:learning_curves}C.5), the higher order mutations were mutant based shuffled (Fig. \ref{fig:learning_curves}C.2) or separately randomly shuffled from the lower order mutations (Fig. \ref{fig:learning_curves}C.4). In such cases, the LCs did not present discontinuities as before but showed acceleration and deceleration. This was also confirmed when we extended the vocabulary of possible mutations (Fig. \ref{appendix:fig:2}) and when we further restricted the mutagenizable positions (Fig. \ref{appendix:fig:3}). The presence or absence of the WT (parent) sequence in the training set also strongly affected the path of the corresponding LCs (see below). 

In order to extend our findings to a different type discrete combinatorial space, we applied the same methodology to a linear molecular graph (hexane, Fig. \ref{fig:learning_curves}D.1) and to a cyclic molecule (cyclohexane, Fig. \ref{fig:learning_curves}D.2). In these scenarios, we observed only saturated decay behavior, independently from the mutations contained in the training set. Furthermore, upon \hl{merging} the two molecular datasets, we observed results that were strikingly akin to those previously obtained \hl{(Fig. \ref{fig:learning_curves}D.3)}.

Finally, we studied the impact of the mathematical encoding of our systems. In the peptide scenarios (Fig. \ref{appendix:fig:4} and Fig. \ref{appendix:fig:5}), we utilized a normalized, PCA-reduced version of the AA index database \cite{VD_Vectorization_02} (Tab. \ref{appendix:tab:aaindex}), and a binned z-score \cite{VD_Vectorization_03} (Tab. \ref{appendix:tab:zscore_binned}). The outcomes were congruent with prior observations, reinforcing the initial findings. In the organic small molecule cases, a \hl{similar} outcome was observed when\hl{ both} Coulomb matrix\hl{-based} encoding \cite{VD_Vectorization_04} \hl{as well as Extended-Connectivity Fingerprints with radius 2 (ECFP4) \cite{VD_Vectorization_05} were} employed (Fig. \ref{appendix:fig:6}\hl{ and \ref{appendix:fig:6_5}}). \hl{In these cases, the learning dynamics shifted from saturation-type decays to asymptotic decays, indicating that, in addition to the nonlinearity of the predicted fitness landscape, the encoding itself could determine the shape of the LCs. This indicates that neither factor could remove the discontinuities observed when applying mutation-based shuffling; instead, they only controlled which decay regime the LCs exhibit.} \hl{Interestingly, and in contrast to what was observed previously for EvoE}, when the entire chemical dataset was randomly shuffled, neither acceleration nor deceleration was observed. This phenomenon\hl{, which might arise due to case specific causes,} warrants additional investigation in future studies. 

\subsection{Learning and predicting peptide fitness with variable amounts of training data} 

\begin{figure}[htb!]
  \centering
  \includegraphics[width=\textwidth]{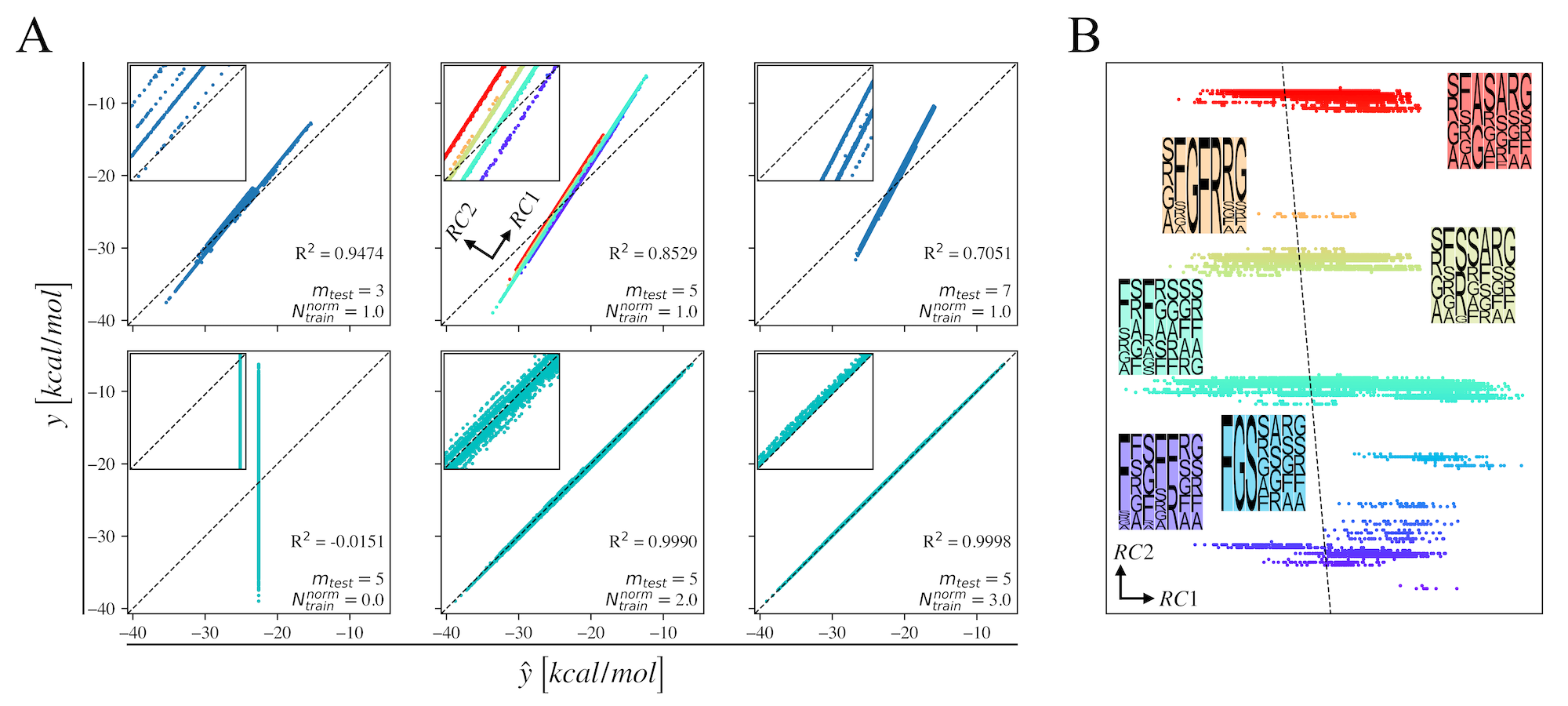}
  \caption[Calibration plots of Fg-$\beta$ / S. epidermidis adhesin SdrG complex binding energy function.]{Calibration plots of Fg-$\beta$ / S. epidermidis adhesin SdrG complex binding energy function. (A) Scatter plots showing true ($y$) vs. predicted ($\hat{y}$) energies at different numbers of mutations ($m$, first row) and training instances ($N_{training}^{norm}$, second row). Insets: zoom-in. (B) Rotation of the predictions coming from a ML model trained with the WT and all single mutants and tested on all quintuple mutants (shown in panel A, 1st row, 2nd column). Insets: amino acids frequencies accordingly to their cluster positions.}
  \label{fig:calibration_plots}
\end{figure}

For the case study in learning EvoEF binding energy of the Fg-$\beta$ peptide, we next generated calibration plots (Fig. \ref{fig:calibration_plots}A), which consisted of plotting the EvoEF binding fitness values against the results predicted by the ML model. We achieved this by separating the data according to the number of mutations enclosed (Fig. \ref{fig:calibration_plots}A, first row and Fig. \ref{appendix:fig:8}, column-wise) and by including different information levels in the training set (Fig. \ref{fig:calibration_plots}A, second row and Fig. \ref{appendix:fig:8}, row-wise). This was done to relate to a realistic scenario where the aim was to explore regions of the search space containing higher number of mutations in comparison to the ones available in the experimental data. In the results presented here, two distinct trends were observed. On one hand, the prediction accuracy of the ML models increased when more information (data) was included in the training set (Fig. \ref{fig:calibration_plots}A, second row), as already observed with the LCs. On the other hand, the model performance scaled inversely with the number of mutations contained in the predicted data (Fig. \ref{fig:calibration_plots}A, first row), highlighting the difficulty of pure ML to extrapolate from the training set.

Another observation was that when the ML model was trained with the WT sequence and all single mutants (Fig. \ref{fig:calibration_plots}A, first row, second column), we noticed a distinct pattern arising from the predictions, where the predicted points formed different parallel clusters (Fig. \ref{fig:calibration_plots}B). These clusters exhibited an angular deviation, relative to the line of “perfect prediction” (Predicted = Actual), indicating a systematic deviation from the ideal alignment. Such deviations seemed to be mitigated by the inclusion of additional information (mutations) in the training set, or by predicting data containing a specific number of mutations, following the same trends observed for the model performance. Moreover, the incorporation of supplementary information in the training set also influenced the clusters dispersion, resulting in a reduced separation among them (Fig. \ref{fig:calibration_plots}A, insets). Upon examining the frequency distribution of mutations within the clusters (Fig. \ref{fig:calibration_plots}B, Fig. \ref{appendix:fig:7}), we did not identify any specific trends, except for a notable accumulation of phenylalanine (F) at the first position in the lower clusters.

These latest observations led us to further investigate how the predictions were distributed in comparison to the actual values. This was undertaken with an expanded focus, not solely on the learning points where complete sets of mutations were included during training, as previously done, but also at intermediate stages. By adopting this approach, new trends were observed. At first, we concentrated on the phase where the WT and the single mutants were gradually incorporated in the training set (saturated decay, Fig. \ref{appendix:fig:9}). At the initial stage, where only a limited set of datapoints were included, we observed a superimposing phase (overfitting), in which the test values were simply inferred from the nearest neighbor contained in the training set. This was the natural extension of the scenario where only a single datapoint (the WT) was used for training and a single vertical line was observed in the calibration plots (Fig. \ref{fig:calibration_plots}A, second row, first column). The number of vertical lines identified increased as the training dataset grew, until it reached a point where they could no longer be distinguished from one another. Shortly after, a shift in the learning paradigm appeared to occur, leading to the emergence of different parallel clusters, which no longer resulted vertically aligned plots, but rather tilted plots. Significantly, continuing to add new data points to the training set resulted in the rearranging and merging of such clusters. This behavior persisted until all single mutants, and the WT were incorporated and a minimal count of clusters was achieved. Upon incorporating the double mutants (asymptotical decay, Fig. \ref{appendix:fig:10}), the clusters once again diverged and became fuzzier in an initial phase, before converging in a different arrangement during the later stage.

A comparable occurrence was observed when the study was extended to many-body response variables, on both linear (Fig. \ref{appendix:fig:11}, Fig. \ref{appendix:fig:12}) and non-linear (Fig. \ref{appendix:fig:13}, Fig. \ref{appendix:fig:14}) functions. It is important to highlight that in the latter scenario, the clusters that formed did not align linearly but exhibited a non-linear relationship.

\subsection{The impact of specific sequence examples on learning }

\begin{figure}[htb!]
  \centering
  \includegraphics[width=\textwidth]{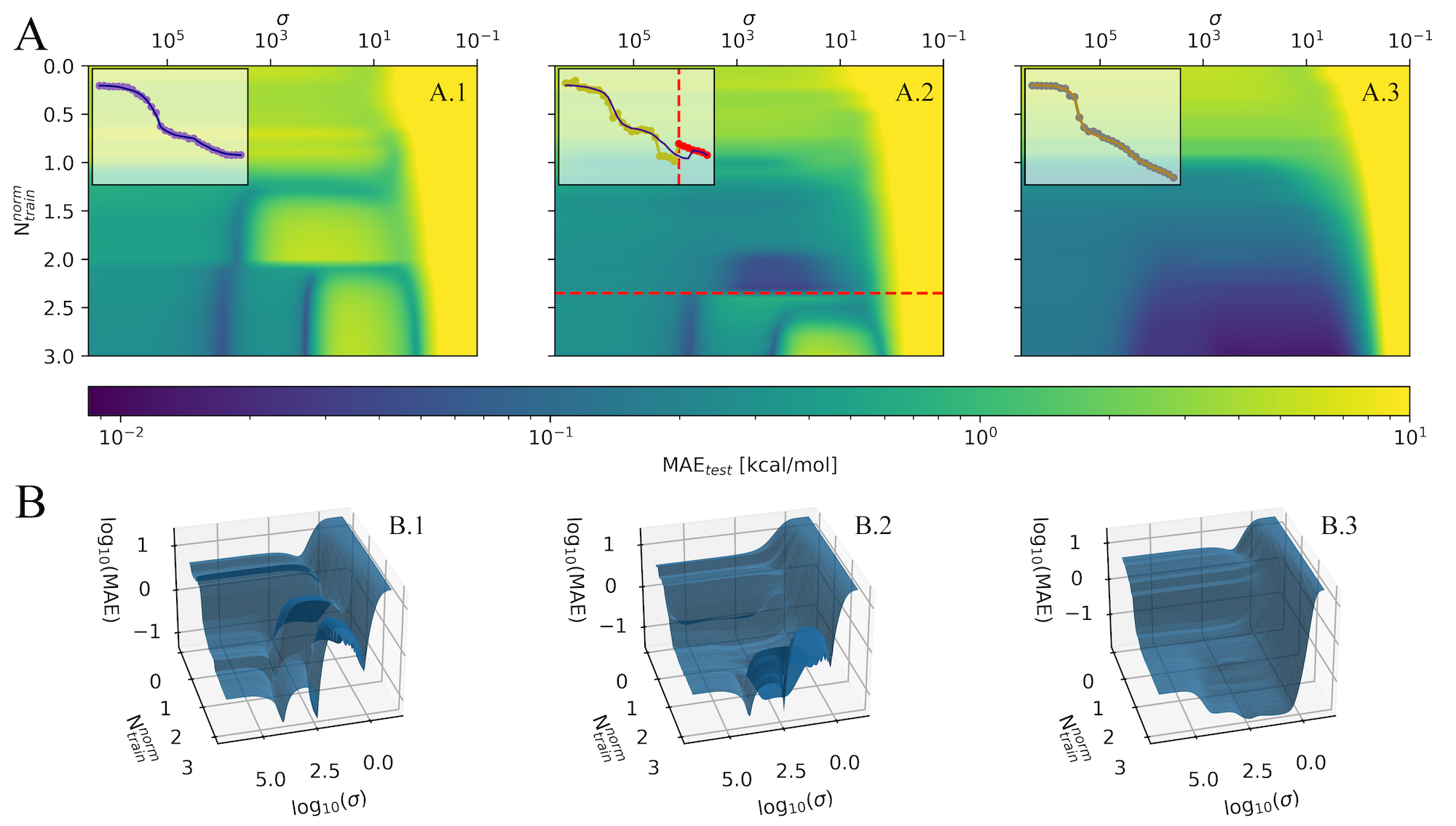}
  \caption[Impact of WT sequence being included in the training data on learning Fg-$\beta$ / SdrG complex binding energy function (EvoEF).]{Impact of WT sequence being included in the training data on learning Fg-$\beta$ / SdrG complex binding energy function (EvoEF). (A) Extended LCs (hyperparameter, $\sigma$) of single examples using different shuffling strategies (see Fig. \ref{fig:workflow_overview}). The dashed red lines mark the point at which the WT sequence was included in the training set. Insets: standard LCs of the specific replicates analysed. (B) 3D projection of the plots shown in panel A.}
  \label{fig:WT_impact}
\end{figure}

\hl{F}urther investigations were conducted to expand upon the preliminary observation presented above regarding the impact of including the WT on the LCs in cases where the training set order was randomized (red paths, Fig. \ref{fig:learning_curves}C.2, \ref{fig:learning_curves}C.4 and \ref{fig:learning_curves}C.5). To achieve this, we extended the LC of three specific examples by calculating the test MAE on each kernel scale value used during the hyperparameter optimization (Fig. \ref{fig:WT_impact}). On one hand, when the training set was mutant based shuffled and the remaining mutants were randomly shuffled (Fig. \ref{fig:learning_curves}C.3), steep valleys arose (Fig. \ref{fig:WT_impact}A.1 and Fig. \ref{fig:WT_impact}B.1). The number of such valleys seemed to correlate with the number of mutations included in the mutant based shuffled training dataset (i.e., one for single mutants, two for doubles, etc.). Moreover, the sudden change in the order of magnitude of the optimal hyperparameter, due to a jump from a valley to another one, could partially explain the learning discontinuities observed within this work. On the other hand, when the whole dataset was shuffled (Fig. \ref{fig:learning_curves}C.5), a trough-like configuration replaced the previous scenario (Fig. \ref{fig:WT_impact}A.3 and Fig. \ref{fig:WT_impact}B.3). In this case, the optimal path would follow a gentle, sloping descent down the hills, minimizing steep declines and allowing for a smoother transition to the valley floor, explaining the learning accelerations/decelerations previously observed. The last example showed a situation where the WT sequence was included in the training set in the middle of the LC (Fig. \ref{fig:WT_impact}A.2 and Fig. \ref{fig:WT_impact}B.2).  We already noted that such incorporation significantly affected the LC path, suddenly increasing the test MAE. Interestingly, the extended LC of this example showed a combination of the two behaviors observed above. In fact, before this critical point, the extended LC presented one single, broad opening. Subsequently, the surface transformed immediately, revealing a scenario where multiple valleys became distinctly evident, similar to what was observed in the mutant based shuffled trained example (Fig. \ref{fig:WT_impact}A.1 and Fig. \ref{fig:WT_impact}B.1).

\subsection{\hl{Validation of learning behavior on experimental data}}

\hl{Following our analysis of in-silico data, we further examined learning dynamics using a dataset from a complete mutational scan of four positions in the GB1 protein domain, which provides experimentally measured fitness values reflecting protein stability and binding affinity. This allowed us to assess whether the findings derived from the synthetic benchmarks extend to a real-world experimental setting. To this end, we generated LCs from the GB1 dataset using mutant-based and random shuffling strategies in combination with OHE and ESM embeddings (Fig. \ref{appendix:fig:15}). The mutant-based shuffling reproduced the trends observed with synthetic data, showing clear discontinuities and asymptotic decays across both encoding strategies. In contrast, random shuffling did not yield the multiple accelerations and decelerations seen in the EvoEF case, similarly to what was observed in the molecular case studies using synthetic data. This might be due to the large variability and heteroscedastic noise inherent to real-world DMS measurements. This interpretation was consistent with the substantial variability apparent in the individual learning curves.}

\section{Conclusion}

In this work we studied data-error scaling\hl{ laws} in machine learning in the context of mutagenisable discrete spaces, in particular a peptide interacting with its target receptor (Fg-$\beta$:SdrG)\hl{,} small molecule solvation energies\hl{, and a larger protein (i.e., GB1)}. We \hl{first} leveraged deterministic functions to compute the response variables and generate synthetic datasets. This approach enabled us to focus on a learning process devoid of noise, effectively eliminating variability attributed to experimental errors or non-deterministic simulations. \hl{We then validated the trends identified in these controlled synthetic benchmarks using a real-world dataset from an experimental DMS campaign.}

We presented a novel way to normalize LCs based on mutational saturation of a particular degree, and a new shuffling methodology suitable for any discrete combinatorial input space prone-to-mutations. Our findings indicated that the learning process did not conform to a single power law but might be discontinuous, and categorizable in two distinct regimes (i.e. asymptotical or saturated decay). These regimes exhibit acceleration and deceleration phases of learning. Such knowledge could be extremely valuable in Design Of Experiments (DoE), as well as in developing theoretical mechanistic models. In the first scenario, given knowledge on the order of the phenomenon involved, a lab scientist could design a variant library to saturate certain numbers of mutations (saturated decay) and to partially cover other ones (asymptotic decay). In the second scenario, analyzing the trajectories of LCs could provide significant insights into the order of the model (i.e., linear or non-linear) that would be appropriate for approximating a physico-chemical phenomena. Based on our results, we showed that learning trajectories can be affected by the shuffling strategy applied to the data, by the encoding strategies employed, and by the natural process leading to the response variable. Furthermore, our analysis demonstrated that test predictions clustered in parallel groups and that the presence of the parent sequence (i.e. WT) from which variants are derived through mutations can dramatically influence the learning process.

Although our findings appear to diverge from previous observations in the domains of theoretical chemistry and physics, this discrepancy may be attributed to the nature of datasets commonly utilized within these fields. Chemistry, as a discipline, focuses predominantly on the de novo design of molecules, prioritizing alterations in molecular graphs over introducing point mutations on existing structures. This approach contrasts with the methodologies typically applied to generate biological DMS data. In fact, most of the potential molecular variants would be highly unstable, which diminishes their practical applicability and, consequently, the incentive to incorporate them into simulations for data generation. Despite this, our results hold value for researchers in molecular disciplines where mutations are feasible. For instance, they could inform studies investigating the effects of varying functional groups at specific positions within a molecule. 

Given the complexity of understanding the data-error \hl{scaling laws }in the context of discrete input spaces prone-to-mutations, this study \hl{mainly} focused on noise-free labelled data on point mutated molecular and peptide graphs\hl{, while also incorporating results from an experimental dataset to validate our findings}. Future studies could extend our work by \hl{introducing} \hl{controlled noise} in the response variable(s) \hl{to study the role of heteroscedastic uncertainty in DMS experiments, and by investigating why accelerations and decelerations may disappear under some circumstances when using random shuffling}. Our findings on data-error \hl{scaling laws } could impact the way experimental and simulation campaigns are conducted for mutational data, lowering both the time and the costs of potentially any field where discrete combinatorial inputs prone-to-mutations play a role. \hl{Our contribution is to highlight that exotic learning behaviors (i.e., saturated or asymptotic decay in mutant-based shuffled problems) exist and can guide the design of the initial mutational library itself in DMS campaigns. For instance, if a property of interest is expected to show asymptotic decay at a mutational level (e.g., double mutants), fully saturating that level would be unnecessary, and resources would be better allocated elsewhere. Conversely, if saturated decay arises at a mutational level, ensuring full coverage there would be more efficient.} Furthermore, our findings have the potential to broaden the statistical learning theory, contributing to the understanding of how machines learn from specific \hl{combinatorial} data.

\begin{ack}
This work was supported by the University of Basel, the Swiss Federal Institute of Technology in Zürich (ETH Zürich) and the Swiss Nanoscience Institute (SNI, project P1802). The authors declare that some text was edited using ChatGPT (\href{https://chat.openai.com}{https://chat.openai.com}). The editing focused solely on rearranging, enhancing, or correcting the syntax of existing sentences without creating new content. Some of the calculations were performed at sciCORE scientific computing center at the University of Basel (\href{https://scicore.unibas.ch}{https://scicore.unibas.ch}). 3D molecular structures were generated and processed with Leruli \cite{LeRuLi} and PyMOL \cite{PyMOL}. QMLCode \cite{QMLCode} was utilized for some computations. The authors would like to thank Dr. Dominik Lemm and Dr. Stefan Heinen for the insightful discussions in the preliminary phase of this project.
\end{ack}


\putbib 
\end{bibunit}


\newpage
\appendix

\makeatletter
\providecommand{\@titlecalledfalse}{} 
\@titlecalledfalse

\def\@title{Supporting Information\\Data-Error Scaling\hl{ Laws} in Machine Learning on Combinatorial Mutation-prone Sets: \hl{Proteins }and Small Molecules}
\def\@author{Vanni Doffini$^{1,2,3}$ \quad O. Anatole Von Lilienfeld$^{4,5,6}$  \quad Michael A. Nash$^{1,2,3}$\\
$^1$\textit{University of Basel} \quad
$^2$\textit{ETH Zurich} \quad
$^3$\textit{Swiss Nanoscience Institute} \\
$^4$\textit{University of Toronto}\quad
$^5$\textit{Vector Institute}\quad
$^6$\textit{TU Berlin}}
\def\@date{}

\@maketitle
\makeatother

%

\renewcommand\thefigure{S\arabic{figure}}    
\setcounter{figure}{0}  

\renewcommand\thetable{S\arabic{table}}    
\setcounter{table}{0}  

\begin{bibunit}

\section{Code and Data Availability Statement}
The data and code produced for this work, including the documentation associated, are available free of charge on \href{https://doi.org/10.5281/zenodo.11148308}{10.5281/zenodo.11148308} and \href{https://github.com/Nash-Lab/DiscontinuousLCs}{https://github.com/Nash-Lab/DiscontinuousLCs}.

\section{Materials and Methods}
\subsection{\hl{Synthetic Dataset} Generation}

We calculated the number of specific and total possible mutations in a discretised space with fixed structure with equations (\ref{eq:n_mut}) and  (\ref{eq:n_cumulative_mut}). 

\begin{equation}\label{eq:n_mut}
D^{disc}_{spec}(n,p,v)=\binom{n}{p}\cdot(v-1)^p 
\end{equation}

\begin{equation}\label{eq:n_cumulative_mut}
D_{tot}^{disc}(n,m,v)= \sum_{p=0}^{m} \binom{n}{p}\cdot(v-1)^p 
\end{equation}

Where $n$ is the number of positions allowed to mutate, $p$ and $m$ are the number of mutations, and $v$ is the vocabulary size, i.e., the number of possible building blocks that can be drawn. It should be noted that these equations hold true exclusively under the condition that all constituent building blocks (whether amino acids or atoms) of the initial sequence (WT) are included in the vocabulary. Should a scenario arise where none of these components are present, a simplification of the equations can be achieved by discarding the “$-1$” term. For scenarios that fall between these extremes, it becomes necessary to formulate more specialized and less intuitive expressions tailored to each specific case. The equations in question were utilized to evaluate the feasibility of creating in-silico libraries encompassing both a biological and a chemical system.
For the biological system, we chose a segment of the modified Fg-$\beta$ peptide (\textit{FFFSARG}, $n=7$) to be the WT (Fig. \ref{fig:workflow_overview}A.1). This specific sequence was previously characterized by 
\textcite{VD_SdrGFgB_1r17}.
During the main study, we limited the vocabulary to the amino acids already present in the initial sequence ($v=5$), namely: Alanine (A), Phenylalanine (F), Glycine (G), Arginine (R) and Serine (S). This allowed us to generate an unlabelled database of possible mutations (Fig. \ref{fig:workflow_overview}A.2). Furthermore, to enhance the robustness of our study, we expanded our analysis to incorporate Cysteine (C), due to the inclusion of a sulfur atom; Aspartic acid (D), attributed to its negatively charged side chain; and Tryptophan (W), which is distinguished as the largest and heaviest amino acid. 
For the chemical system, the selection was made on hexane (C$_{6}$H$_{14}$, $n=6$) and cyclohexane (C$_{6}$H$_{12}$, $n=6$), allowing the substitution of the carbon atoms with oxygen(s) or nitrogen(s). The hybridisation was fixed to Sp3 and the structures were filled accordingly with hydrogen atoms. It should be noted that, given molecules typically exhibit linear or nonlinear graph-based structures lacking a clear direction of interpretation, certain mutations were functionally equivalent due to symmetries. However, for simplicity, these equivalences were disregarded in our study.

A summary of the number of mutants is provided in Supporting Table \ref{tab:DatabaseMutations}.

We employed different functions to label our generated databases. In the peptide case, we started with two functions inspired by the many-body theory from quantum physics \cite{Book_ML_04}. To maintain the initial analysis as a proof of concept, we implemented the following simplifications: (i) amino acids were utilized as the primary units for computation, thereby avoiding a reduction to the atomic level; (ii) the spatial separation between distinct units was quantified using the integer difference in their positions, as described in equation (\ref{eq:2body_distance}); and (iii) the hash maps  $g_1\{\cdot\}$ and $g_2\{\cdot\}$ (Tab. \ref{tab:Dictionaries}), assigning a unique number to each amino acid in the vocabulary, were established through random sampling rather than derived from physico-chemical properties. This last approach was chosen to preclude any potential bias that might arise from prioritizing one chemical/physical attribute over another. The mechanisms underlying one-body and two-body terms interactions were encapsulated by equations (\ref{eq:one_body}) and (\ref{eq:two_body}), respectively.

\begin{equation}\label{eq:one_body}
f_{1body}(x)=\sum_{k = 1}^{L_x} g_1\{x_k\} 
\end{equation}

\begin{equation}
\label{eq:two_body}
f_{2body}(x)=\sum_{k = 1}^{L_x} g_1\{x_k\} + \sum_{i=1}^{L_{x}-1}\sum_{j=i+1}^{L_{x}} \frac{g_2\{x_i\}\cdot g_2\{x_j\}}{d(i,j)}
\end{equation}

Where $x$ is a peptide, $x_{k / i / j}$ is the amino acid of x at position k, i or j, and d is the distance between the different amino acids (eq. \ref{eq:2body_distance}). 

\begin{equation}\label{eq:2body_distance}
d(i,j)=\left| j-i\right|
\end{equation}

Subsequentially, we utilized EvoDesign physcical Energy Function (EvoEF), a composite energy force field developed by the Zhang group \cite{DL_PeptidesBinding_01,DL_PeptidesBinding_02}). Specifically, we used EvoEF1 to estimate the binding energies between the mutagenised Fg-$\beta$ peptides contained in our database and SdrG protein (Algorithm \ref{alg:EvoEF1}and Fig. \ref{fig:workflow_overview}A.3). This corresponded to chains C and A contained in the Protein Data Bank (PDB) database, file 1r17 \cite{VD_SdrGFgB_1r17}, while chains D and B remained unaltered throughout the computational process. To decrease the running time, the loop was divided in 512 instances, which were run in parallel on our computer cluster.

\begin{algorithm}[h]
\caption{EvoEF1 energy calculation}\label{alg:EvoEF1}
\KwData{1r17.pdb}
\KwResult{$energy$}
$1r17\_Repair  \gets RepairStructure(1r17)$\;
\For{$m\leftarrow 0$ \KwTo $M$}{
  $1r17\_Repair\_Mut  \gets BuildMutant(1r17\_Repair)$\;
  $E_{Evo}  \gets ComputeBinding(1r17\_Repair\_Mut\_Repair)$\;
}
\end{algorithm}

Regarding our molecular constructs, we calculated the corresponding solvation energies in water at a fixed temperature of 298 K for each compound contained within the database. To estimate such energies we employed the leruli library \cite{LeRuLi}, which utilized the works of \textcite{LeRuLi_solvation_energy_01} and \textcite{LeRuLi_solvation_energy_02}.

\subsection{\hl{Experimental Data (GB1 Dataset)}}

\hl{In addition to the in-silico datasets described above, we incorporated an experimental dataset derived from DMS of the B1 domain of protein G (GB1), targeting four positions (V39, D40, G41, and V54), as reported by \textcite{VD_DataBio_08}. The dataset comprises fitness measurements, defined as the relative stability and IgG-Fc binding activity of each variant compared to WT, for 149,361 variants (93.4\% of the full combinatorial sequence space), determined experimentally, and 10,639 variants (6.6\%) imputed computationally, as reported in the original study. Together, these measurements provide complete coverage of the entire combinatorial sequence space at the four mutational sites. A breakdown of the variant distribution by mutation count is provided in Supporting Table \ref{tab:DatabaseMutations}.}

\hl{Prior to ML applications, fitness values were log-normalized. To mitigate numerical instability during transformation, zero values were replaced with a small constant (0.0001).}

\subsection{Encoding}

During this study we mainly employed flattened One Hot Encoding (OHE, Fig. \ref{fig:workflow_overview}B) to translate each of our sequences, whether they were constituted by amino acids in a peptide or atoms in a molecule, in a unique, sparse and binary tensor. Even if one had to renounce completely on any physico-chemical information, such representation is still broadly used in the scientific community focused on biological macromolecules and present few advantages. Firstly, the input data could be structured in the same way for both peptides or molecules, making it possible to compare our observations across different topics. Secondly, the process did not require three-dimensional structural data, thus circumventing the need for costly and potentially imprecise simulations  \cite{VD_NeurIPS2023_10,VD_ML4ChemSpace_06}, particularly relevant in the context of peptide-protein interactions. Thirdly, it was not reliant on any pre-trained ML model \cite{VD_NeurIPS2023_03,MN_DataProject1_02,VD_ML4ProtSpace_30} or evolutionary/alignment (MSA) strategy \cite{VD_ML4ProtSpace_38,VD_Attention_02}, rending it well suitable for localized DMS datasets.

When integrating the linear and cyclic molecular graphs, we augmented the OHE matrix by appending an additional column. This column contained a zero for rows corresponding to linear structures and a one for all others, effectively distinguishing between the two molecular configurations.

In addition to OHE, our research was expanded to include a variety of other encoding techniques. For the mutagenized peptides, we utilized a binned version of the z-score, presented by \textcite{VD_Vectorization_03}, and a reduced version of the popular amino acids index database (AA index) \cite{VD_Vectorization_02}. The first alternative encoding, involved categorizing five physico-chemical properties (lipophilicity, size, polarity, electronegativity, and electrophilicity) into bins (Tab. \ref{appendix:tab:zscore_binned}). This process yielded a binary tensor analogus to OHE, with the distinction of retaining some degree of the physico-chemical information. The second method, consisted in compressing the AA index to 18 variables per amino acid through the application of principal component analysis (PCA) for dimensionality reduction (Tab. \ref{appendix:tab:aaindex}). Moreover, each variable was normalized utilizing a min-max scaling approach to ensure uniformity in the encoding scale. Regarding our molecular constructs, they were also represented through encoding with the standard flattened Coulomb matrix \cite{VD_Vectorization_04} (eq. \ref{eq:coulomb_matrix}), in addition to a variation of this matrix that was sorted according to eigenvalues \cite{VD_ML4ChemSpace_25}.

\begin{equation}\label{eq:coulomb_matrix}
    M_{ij} =
    \begin{cases}
        \frac{1}{2}Z_{i}^{2.4} & \text{if }i=j \\
        \\
        \frac{Z_{i}Z_{j}}{{||r_{i}-r_{j}||}_{2}} & \text{if }i\neq j \\
    \end{cases}
\end{equation}

Where $i$ and $j$ are indices of the atoms in the encoded molecule, $Z$ is the nuclear charge and $r$ contains the atomic spatial coordinates.

\hl{In addition to Coulomb matrix–based encodings, molecular constructs were also represented using Extended-Connectivity Fingerprints with radius 2 (ECFP4), a hashed binary vector that encodes circular topological features of molecular substructures \cite{VD_Vectorization_05}. Fingerprints were generated using the RDKit cheminformatics toolkit.}

\hl{For the GB1 dataset, we further complemented standard OHE with embeddings derived from the Evolutionary Scale Modeling (ESM) protein language model \cite{VD_ThesisIntro_098}. Specifically, embeddings were extracted from the final (33rd) hidden layer of the pre-trained ESM-2 model (650M parameters; model ID: esm2\_{\-}t33\_{\-}650M\_{\-}UR50D; repository: {\href{https://github.com/facebookresearch/esm}{github.com/facebookresearch/esm}}). To construct variant sequences, point mutations were introduced into the WT GB1 sequence (M{\-}T{\-}Y{\-}K{\-}L{\-}I{\-}L{\-}N{\-}G{\-}K{\-}T{\-}L{\-}K{\-}G{\-}E{\-}T{\-}T{\-}T{\-}E{\-}A{\-}V{\-}D{\-}A{\-}A{\-}T{\-}A{\-}E{\-}K{\-}V{\-}F{\-}K{\-}Q{\-}Y{\-}A{\-}N{\-}D{\-}N{\-}G{\-}V{\-}D{\-}G{\-}E{\-}W{\-}T{\-}Y{\-}D{\-}D{\-}A{\-}T{\-}K{\-}T{\-}F{\-}T{\-}V{\-}T{\-}E) at positions V39, D40, G41, and/or V54. Each full-length variant sequence was then processed by the model, and residue-level embeddings were obtained. Excluding the special [CLS] and [EOS] tokens at the first and last positions, embeddings were averaged across all remaining residues to produce a fixed-length representation. This procedure yielded a 1,280-dimensional vector per variant, capturing contextual and evolutionary features encoded by the unsupervised training of ESM-2 on large-scale protein sequence databases.}

\subsection{Machine Learning}

Kernel ridge regression (KRR) \cite{Book_ML_12,Book_ML_08,VD_ML4ChemSpace_26,VD_NeurIPS2023_12} was employed as the sole ML algorithm throughout the study (Fig. \ref{fig:workflow_overview}C). One of the primary advantages of this non-parametric method is its closed-form solution (eq. \ref{eq:KRR_closed_form}).

\begin{equation}\label{eq:KRR_closed_form}
\alpha = {\left(K\left(x_{i \in train},x_{j \in train}\right)+\lambda I\right)}^{-1}y_{train}
\end{equation}

Where $K$ denotes the Laplacian shift-invariant kernel (eq. \ref{eq:laplacian_K}). The variables $x_{i/j}$ are the rows associated with the instances of the training set within the flattened OHE matrix (Fig. \ref{fig:workflow_overview}B). The vector $y_{train}$ contains the labels corresponding to the training set. Furthermore, $I$ is the identity matrix and $\lambda$ is the estimation of the label error. For computational stability reasons, $\lambda$ was set equal to $1\cdot 10^{-8}$. The matrix inversion was handled via Cholesky decomposition \cite{VD_Math_01}.

\begin{equation}\label{eq:laplacian_K}
K_{ij}=K\left(x_i,x_j\right)=exp\left(-\frac{{||x_i-x_j||}_{1}}{\sigma}\right)
\end{equation}

The hyperparameter $\sigma$ was optimized by grid search, calculating the MAE on a validation set (eq. \ref{eq:min_valid_error}).

\begin{equation}\label{eq:min_valid_error}
\sigma^{*}=\mathop{\arg \min}\limits_{\sigma} \frac{1}{N_{valid}}\sum_{i\in valid} \left|y_{i} - \sum_{j \in train}K_{ij}\alpha_j\right| 
\end{equation}

Upon determining the optimal scale of the kernel, it could then be applied to predict the response variable for new data points (eq. \ref{eq:inference}).

\begin{equation}\label{eq:inference}
\hat{y}_i=\sum_{j \in train}exp\left(-\frac{{||x_i-x_j||}_1}{\sigma^{*}}\right)\alpha_j \end{equation}

To calculate the Manhattan distance in the Laplacian kernel (eq. \ref{eq:laplacian_K}) we exploited the inherent property of the OHE matrix, which consists exclusively of binary values. This property resulted in the Manhattan distance being equivalent to the square of the Euclidean distance ($d_{e}^2$), thereby enabling its computation through the use of the inner product (eq. \ref{eq:inner_product}).

\begin{equation}\label{eq:inner_product}
{||x_i-x_j||}_{1}=d_{e}^2 = \langle x_i,x_i\rangle + \langle x_j,x_j\rangle - 2\langle x_i,x_j\rangle
\end{equation}

In this study, two distinct strategies were employed for kernel shuffling (Fig. \ref{appendix:fig:1}). The first approach, called random-based shuffling, involved the creation of an index vector with a dimension identical to that of the entire kernel matrix (encompassing all data points). This vector was subsequently randomly shuffled and utilized to reorder both the rows and columns of the kernel matrix accordingly. The second strategy, referred as mutant-based shuffling, also utilized a randomly shuffled index vector, similar to the first method. However, prior to employing this vector for reordering the rows and columns, the indices were rearranged based on the mutations present in the corresponding data points. This approach effectively randomizes the positioning of data points within groups defined by a specific number of mutations. For the implications of shuffling on the composition of training, validation and test set, refer to the following section.

\subsection{Evaluation}

To assess the performance of our models, we employed two methodologies: the generation of calibration plots and the analysis of learning curves (LCs, Fig. \ref{fig:workflow_overview}D). Both strategies were applied to independent test sets. The first technique involved plotting the predicted fitness values ($\hat{y}$) against the true (or "actual") fitness values ($y$), and it was exclusively applied to evaluate the EvoEF case. For the LCs, we developed a new scaling strategy to enclose the combinatorial nature of our sets (Fig. \ref{fig:combinatorial_space_scale}). This was based on the idea that, data enclosing different numbers of mutations provided different amount of information to the ML model. To establish our novel scale, it was imperative to transcend the discrete boundary imposed by equations \ref{eq:n_mut} and \ref{eq:n_cumulative_mut}, allowing for inputs coming from the set of positive real numbers (eq. \ref{eq:cont_cumulative_mut}).

\begin{equation}\label{eq:cont_cumulative_mut}
D_{tot}^{cont}\left(n,\hat{m},v\right)=\sum_{p\in S(\hat{m})}{\binom{n}{p}}_{\Gamma(\cdot)}(v-1)^p 
\end{equation}

This was achieved by substituting the factorial operator by the gamma function in the binomial coefficient (eq. \ref{eq:cont_binomial}).

\begin{equation}\label{eq:cont_binomial}
{\binom{n}{p}}_{\Gamma(\cdot)} = \frac{\Gamma(n+1)}{\Gamma(p+1)\cdot \Gamma(n-p+1)}
\end{equation}

Moreover, $p$ is defined over the set $S$ (eq. \ref{eq:set_p_of_S}), which depends on $\hat{m}$. It is noteworthy that $\hat{m}$ can now assume any real, positive value that is equal to or less than $n$, overcoming the limit of equations \ref{eq:n_mut} and \ref{eq:n_cumulative_mut}.

\begin{equation}\label{eq:set_p_of_S}
S = \left\{ p \in \mathbb{R} \,\middle|\, p = \hat{m}-i,\ i\in \mathbb{N}_{0},\ i\leq \hat{m}+1 \right\}
\end{equation}

Finally, equation \ref{eq:cont_cumulative_mut} can be applied to normalize an arbitrary number of data points ($N$) by addressing an optimization problem (eq. \ref{eq:N_norm}) to derive a normalized variant of it ($N_{norm}$), which was ultimately used in the LCs.

\begin{equation}\label{eq:N_norm}
N^{norm} = \mathop{\arg \min}\limits_{\hat{m} \in [0, n] \subset \mathbb{R}} \left(N - D_{tot}^{cont}\left(n,\hat{m},v\right)\right)
\end{equation}

\begin{figure}[h]
  \centering
  \includegraphics[width=\textwidth]{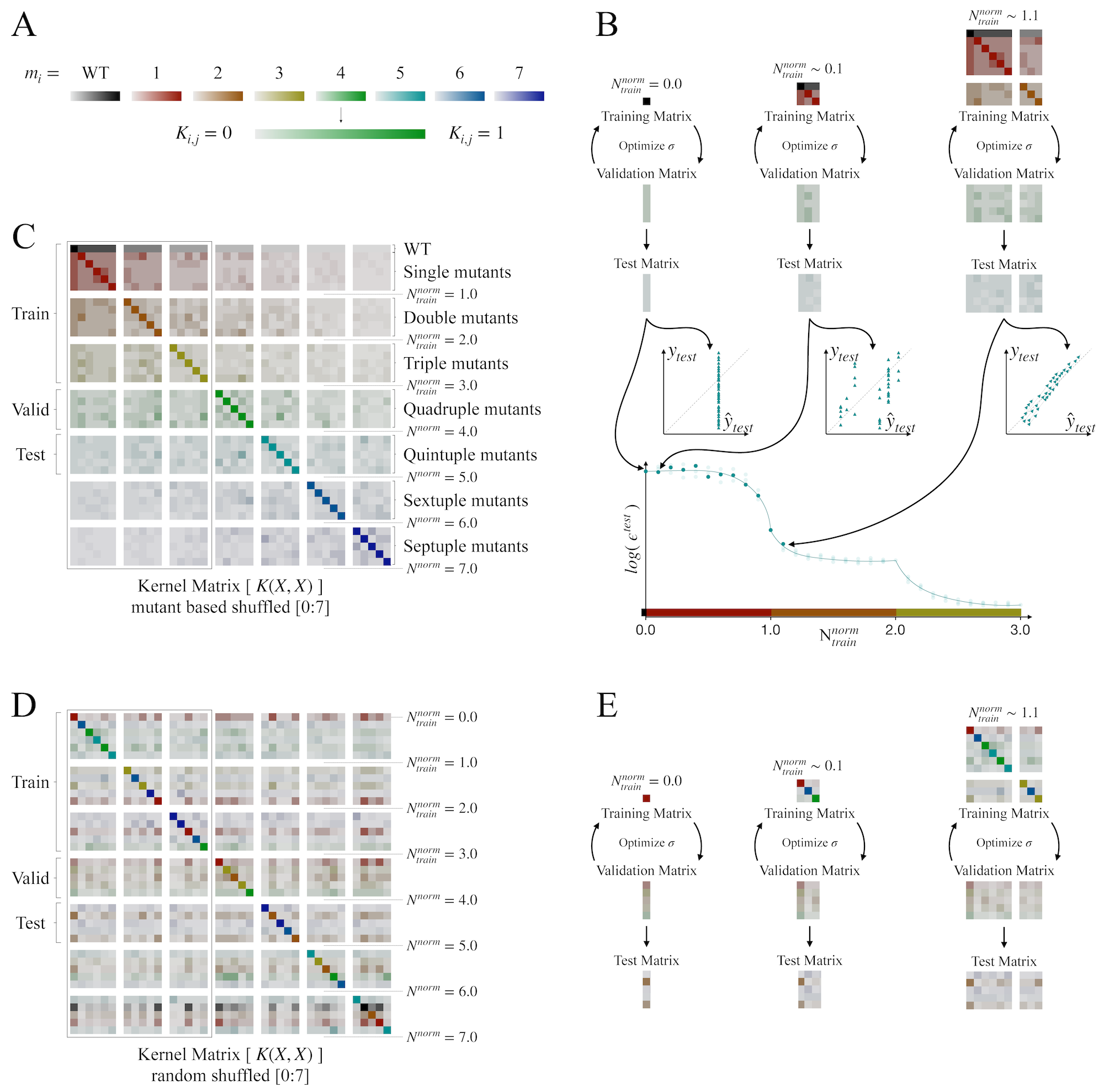}
  \caption[Extension on kernel shuffling techniques.]{Extension on kernel shuffling techniques. (A) Color scheme (row wise): each variant with a specific number of mutations was colored accordingly. The color gradient (light grey – color) follows the specific kernel value (0 - 1). (B) Three examples of training – validation – test pipeline, including LCs and calibration plots, of fully mutant based shuffled kernel (C). (E) Three examples of training – validation – test pipeline of fully random shuffled kernel (D).}
  \label{appendix:fig:1}
\end{figure}

In this research, $N^{norm}$ was utilized to index the rows of the Laplacian kernel derived from the entire dataset. Unless explicitly indicated, the training set was delineated as $N^{norm}$ values ranging from 0.0 to 3.0, the validation set encompassed all data points with $N^{norm}\in (3.0,4.0]$, and the test set included those within $N^{norm}\in (4.0,5.0]$. The choice of shuffling strategy—mutant-based versus randomly-based shuffling—influenced the potential limitation on the dataset composition to a certain subset of mutation numbers (Fig. \ref{appendix:fig:1}). Specifically, employing mutant-based shuffling meant that the validation and test sets were exclusively comprised of quadruple ($m=4$) and quintuple ($m=5$) mutants, respectively. Conversely, with full random shuffling, each subset (training, validation, and test) could contain any mutation number.

\clearpage
\section{Supporting Figures}

\vspace*{\fill}

\vspace*{\fill}

\clearpage

\begin{figure}[p]
  \centering
  \includegraphics[width=\textwidth]{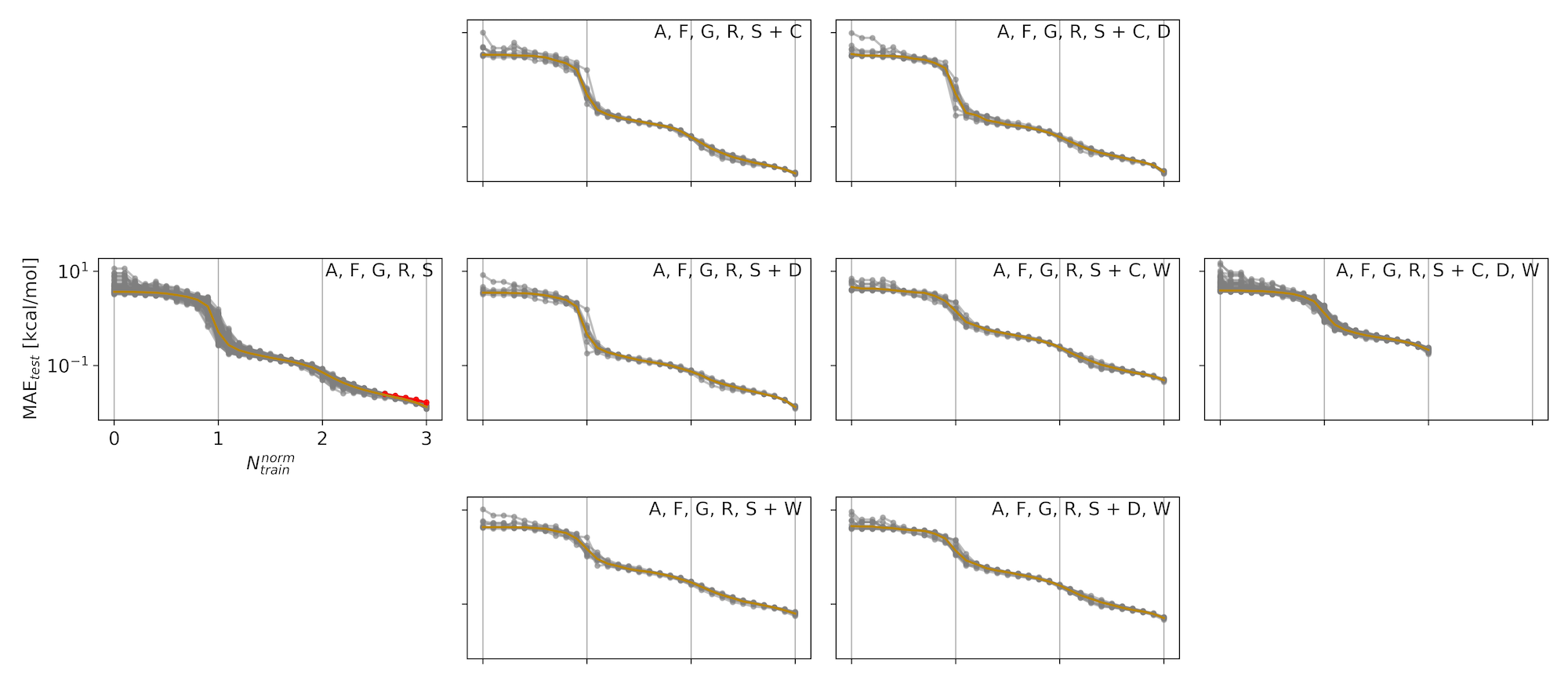}
  \caption[LCs predicting data generated with EvoEF function with variable vocabulary]{LCs predicting data generated with EvoEF function with variable vocabulary (black dots). The random-based shuffling was used across the whole dataset.}
  \label{appendix:fig:2}
\end{figure}

\clearpage

\begin{figure}[p]
  \centering
  \includegraphics[width=\textwidth]{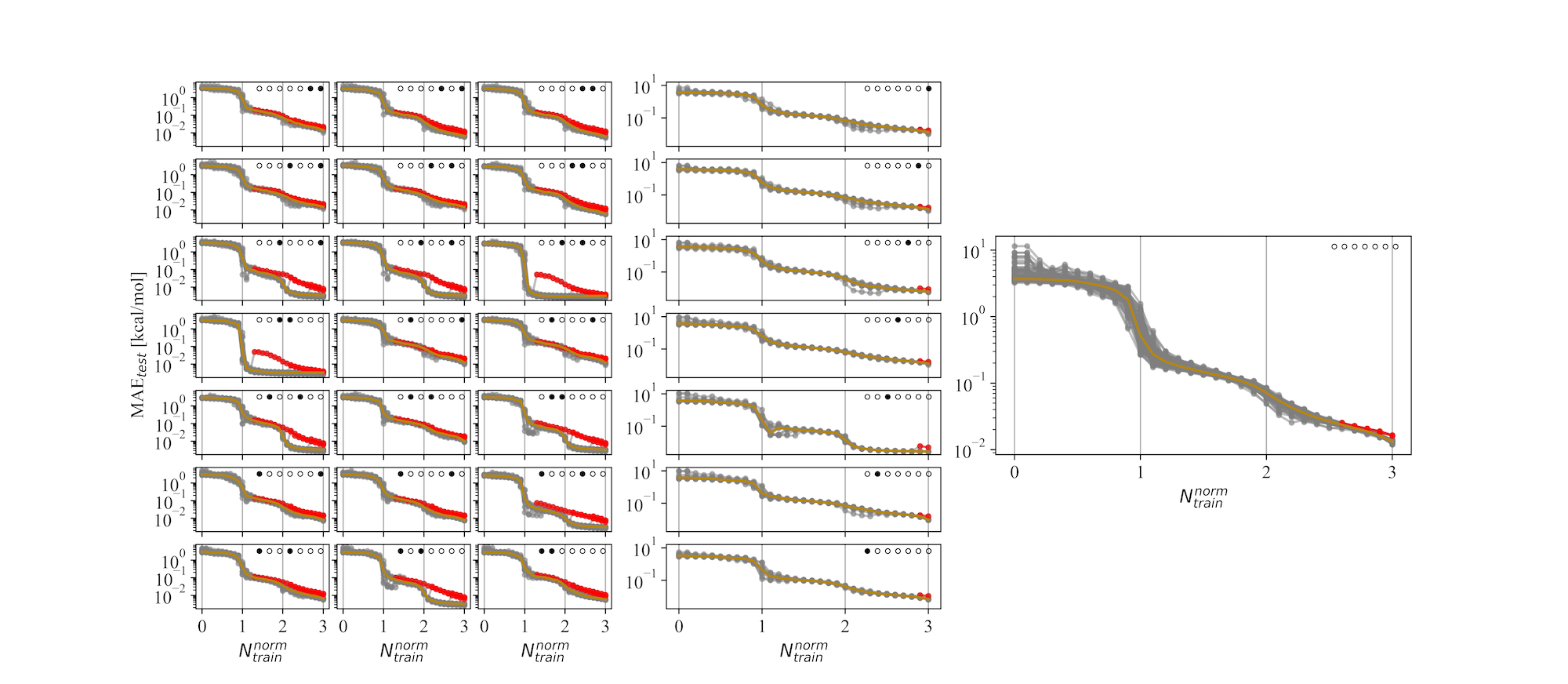}
  \caption[LCs predicting data generated with EvoEF function with unmutagenised positions]{LCs predicting data generated with EvoEF function with unmutagenised positions (black dots). The left side contains all combinations with 5 variable sites (n=7), the central panel contains all combinations with 6 variable sites (n=7) and the right side shows the data reported in the main text. The random-based shuffling was used across the whole dataset.}
  \label{appendix:fig:3}
\end{figure}

\clearpage

\begin{figure}[p]
  \centering
  \includegraphics[width=\textwidth]{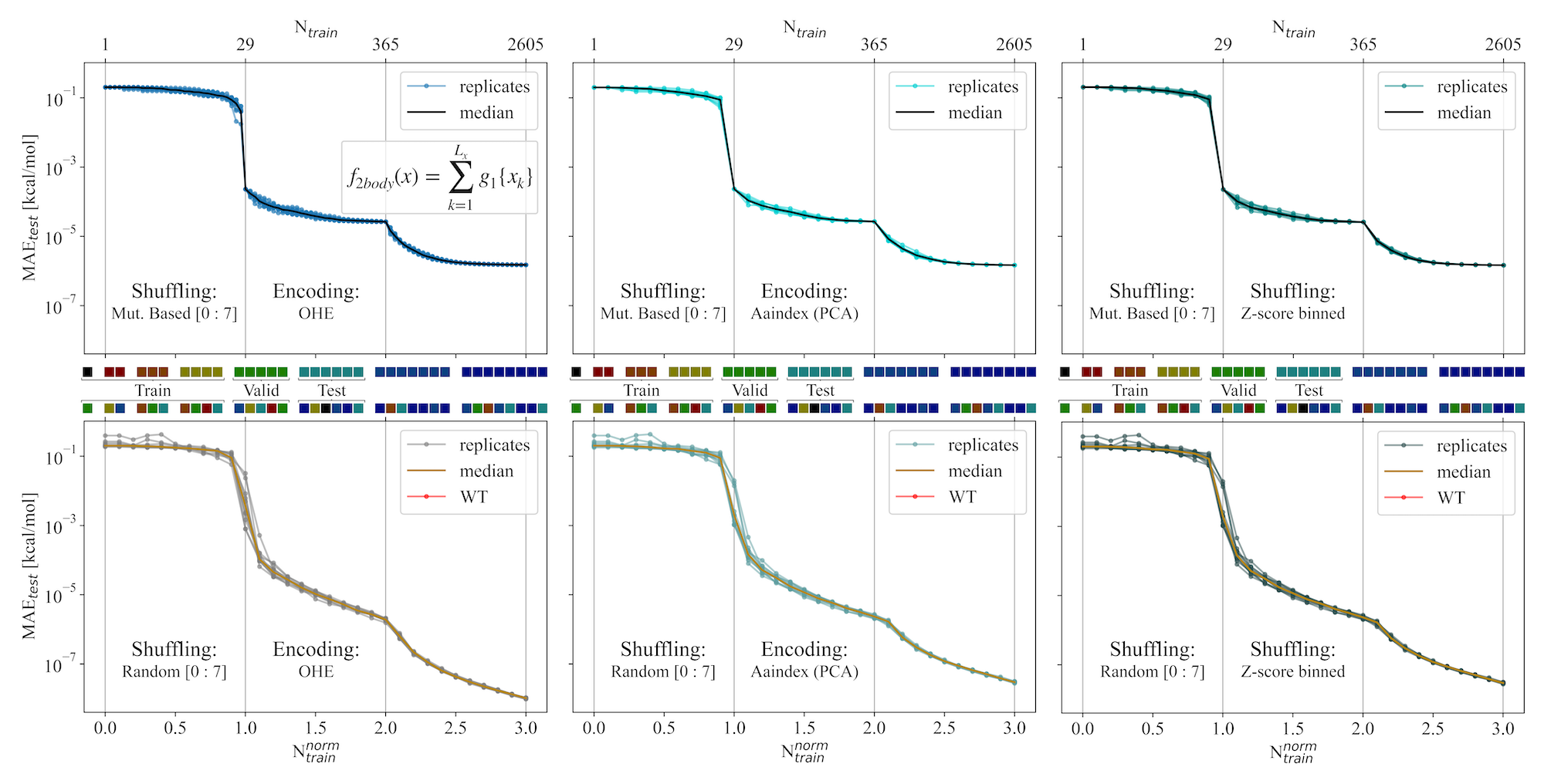}
  \caption{LCs predicting data generated with 1-body-term function with different shuffling (rows) and encoding (columns).}
  \label{appendix:fig:4}
\end{figure}

\clearpage

\begin{figure}[p]
  \centering
  \includegraphics[width=\textwidth]{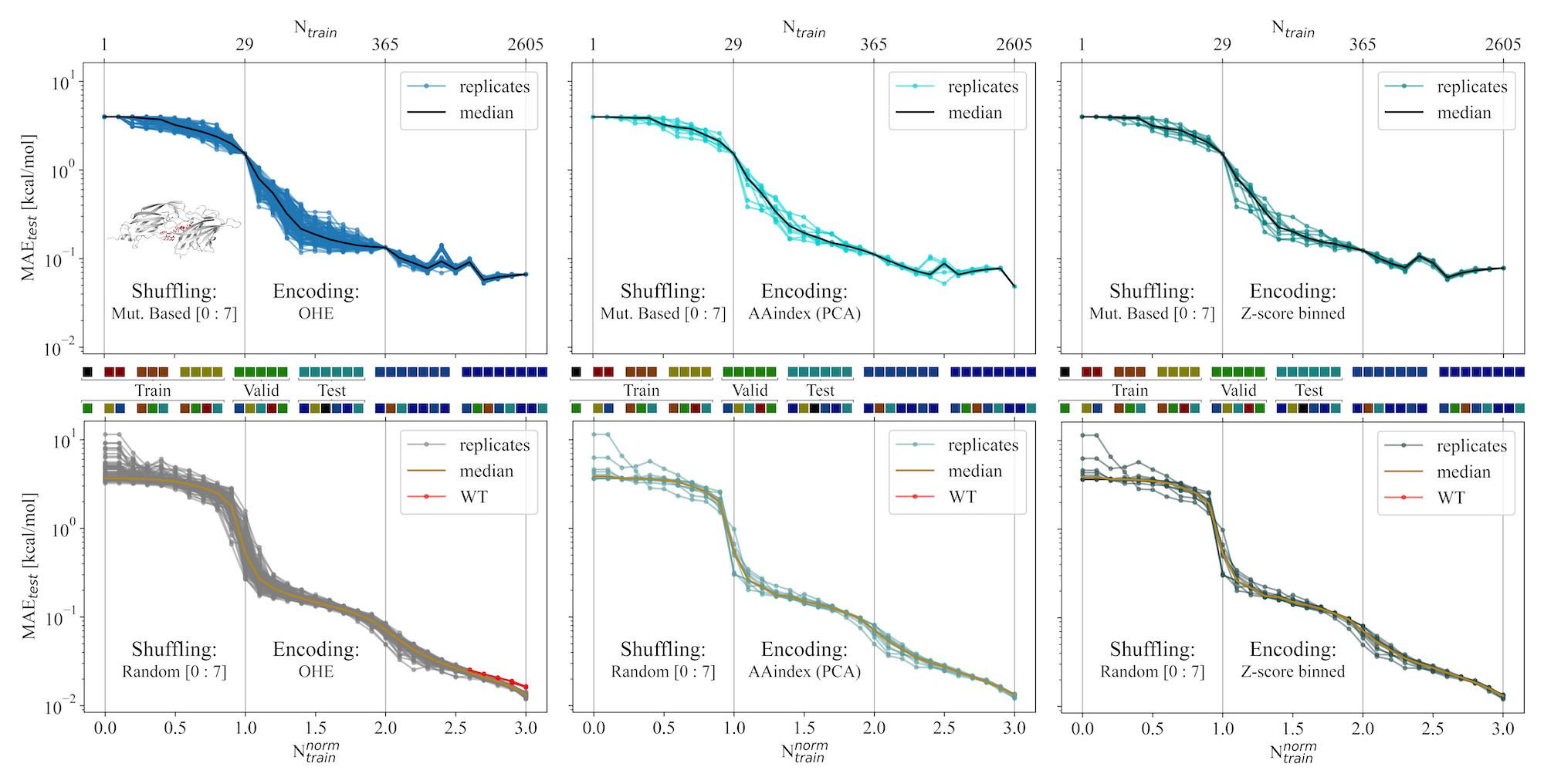}
  \caption{LCs predicting data generated with EvoEF function with different shuffling (rows) and encoding (columns\hl{, from left to right: one hot encoding, PCA-reduced AAindex, and z-score binned}).}
  \label{appendix:fig:5}
\end{figure}

\clearpage

\begin{figure}[p]
  \centering
  \includegraphics[width=\textwidth]{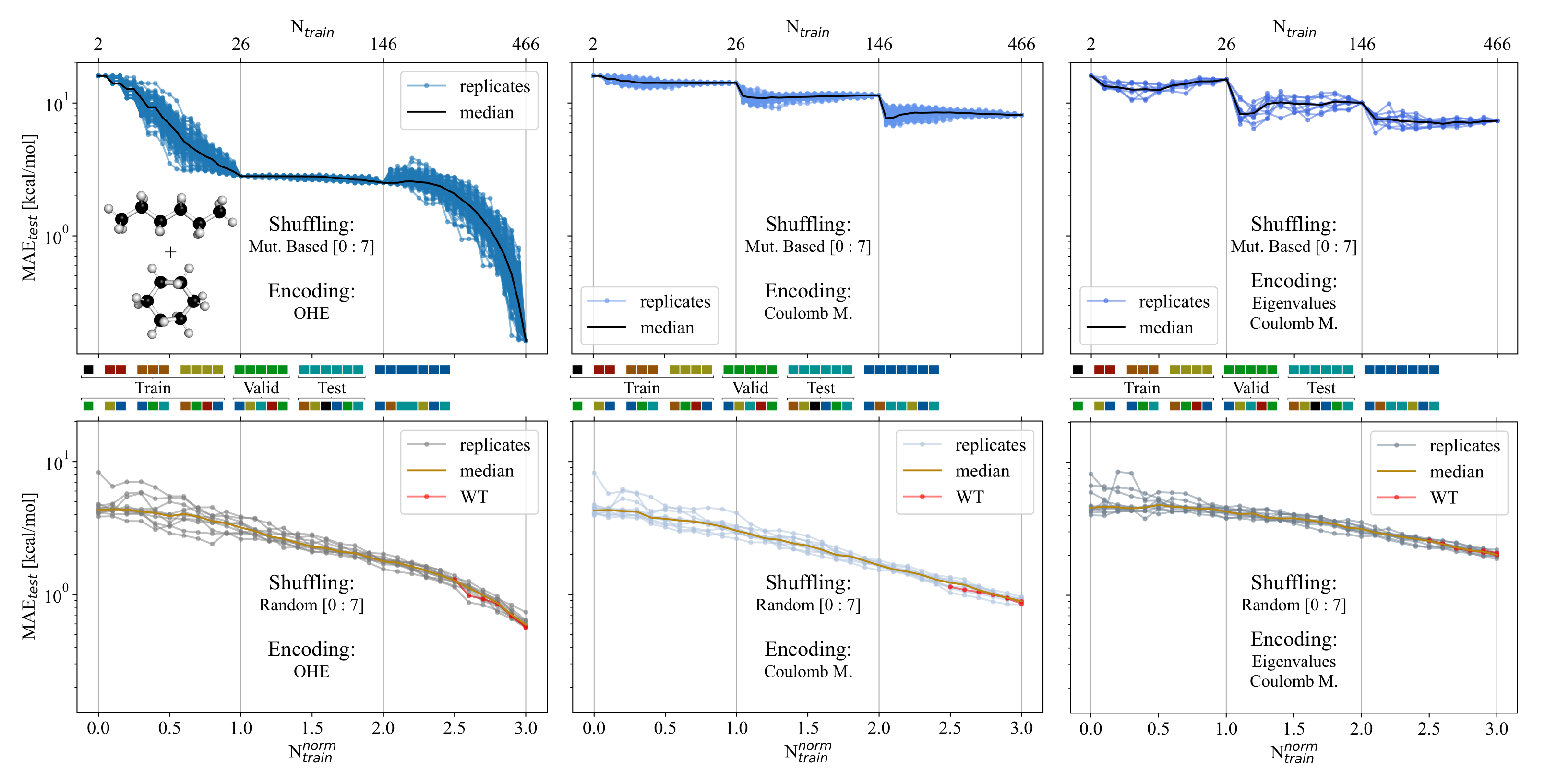}
  \caption{LCs predicting free solvation energies of the molecular database containing the linear (WT = hexane) and cyclic (WT = cyclohexane) graphs with different shuffling (rows) and encoding (columns\hl{, from left to right: one hot encoding, Coulomb matrix, and eigenvalues-sorted Coulomb matrix}).}
  \label{appendix:fig:6}
\end{figure}

\clearpage

\begin{figure}[p]
  \centering
  \includegraphics[width=\textwidth]{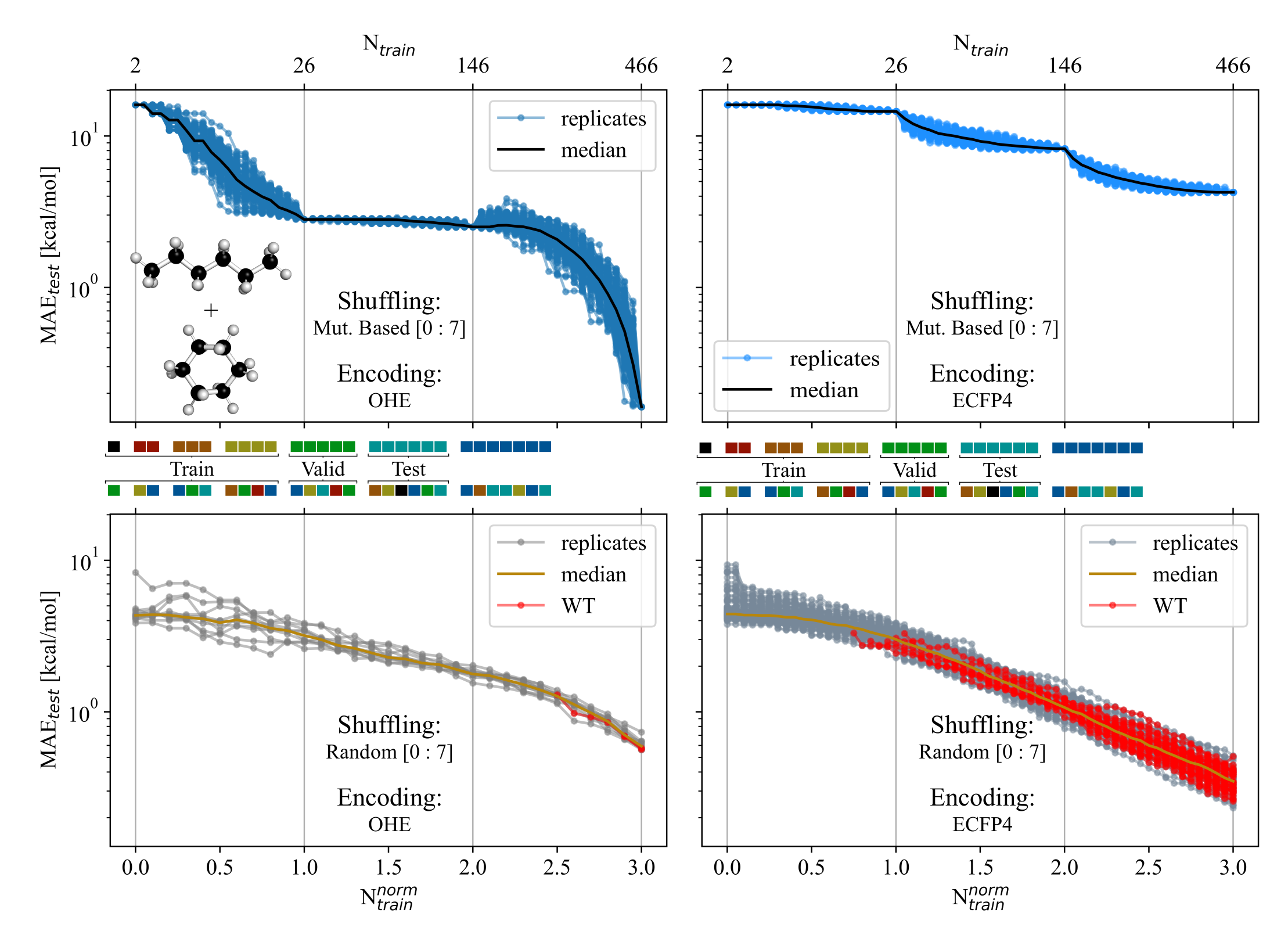}
  \caption{\hl{LCs predicting free solvation energies of the molecular database containing the linear (WT = hexane) and cyclic (WT = cyclohexane) graphs with different shuffling (rows) and encoding (columns, from left to right: one hot encoding, and extended connectivity fingerprint with diameter 4). The OHE subfigures are repeated from Fig. \ref{appendix:fig:6} to facilitate the comparison.}}
  \label{appendix:fig:6_5}
\end{figure}

\clearpage

\begin{figure}[p]
  \centering
  \includegraphics[width=\textwidth]{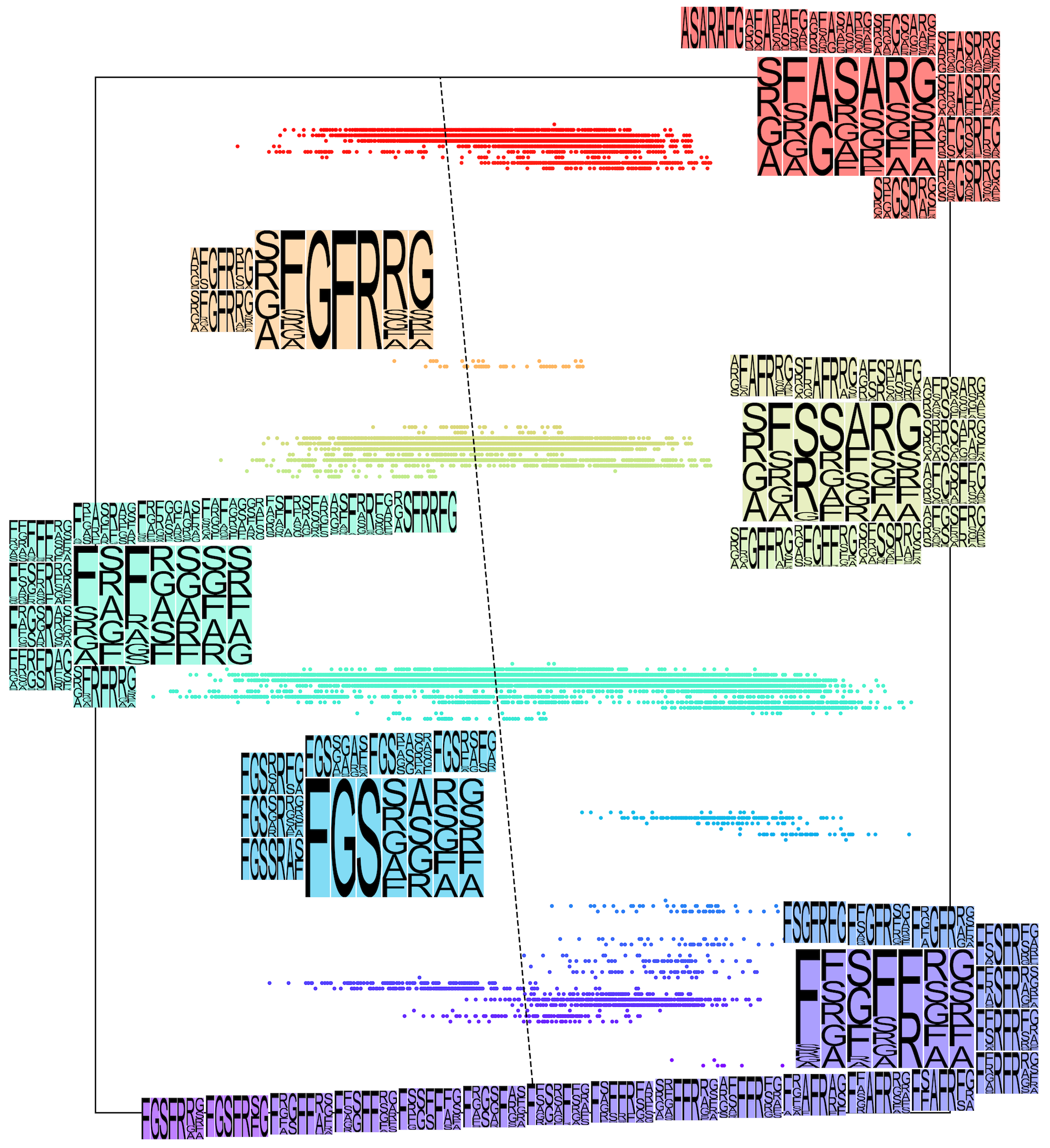}
  \caption{Extension on frequency distribution of amino acids when WT and all single mutants were included in the training set and all quituple mutants were in the test set (Fig. \ref{fig:calibration_plots}B). The frequency distributions of all horizontal clusters were included as insets.}
  \label{appendix:fig:7}
\end{figure}

\clearpage

\begin{figure}[p]
  \centering
  \includegraphics[width=\textwidth]{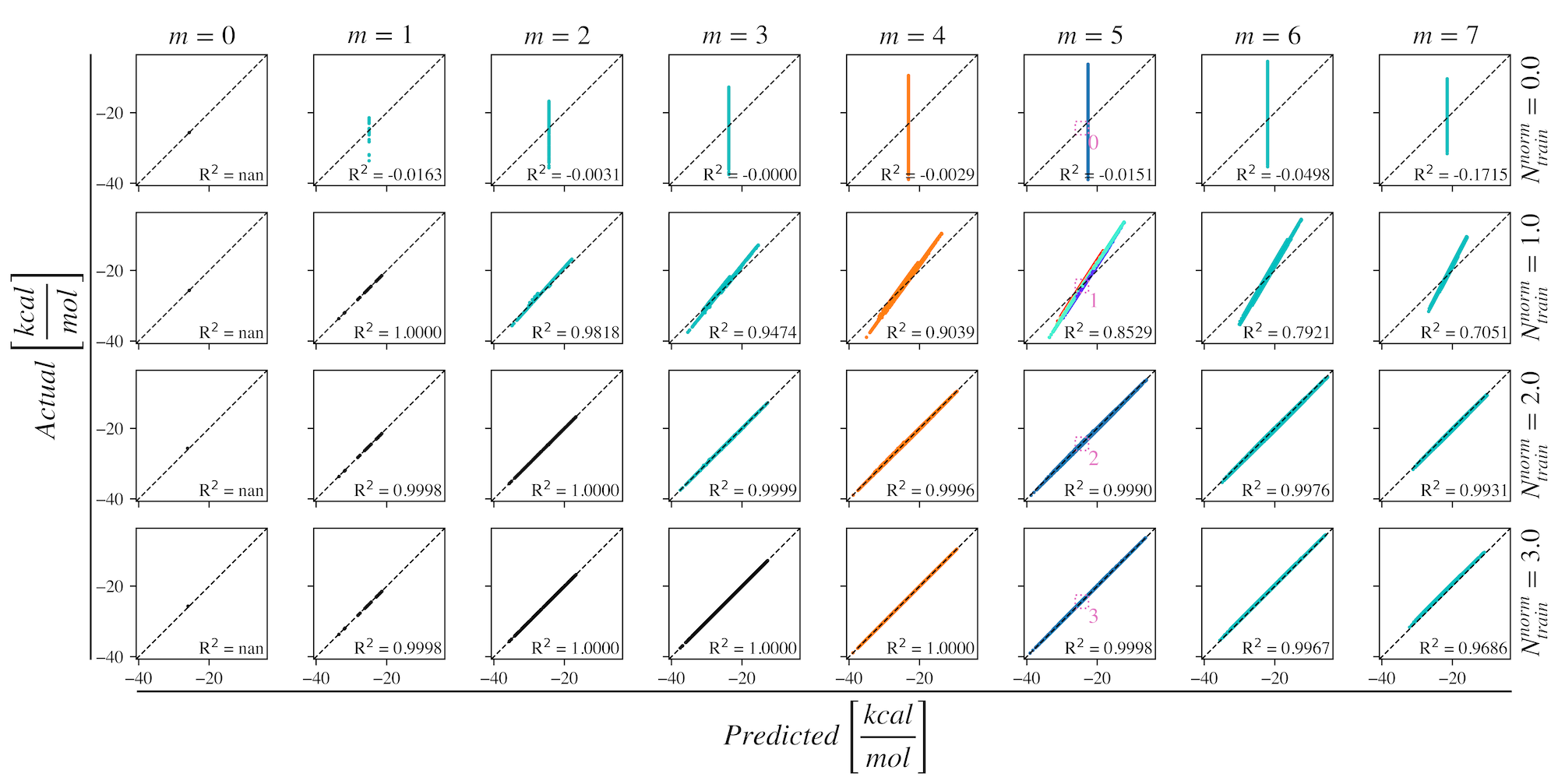}
  \caption{Extension on calibration plots of Fg-$\beta$ / S. epidermidis adhesin SdrG complex binding energy function (EvoEF). Scatter plots showing actual vs. predicted energies. The columns correspond to different numbers of mutations ($m$) while the rows correspond to the number of training examples provided ($N_{train}^{norm}=0.0$ - WT, $N_{train}^{norm}=1.0$ - WT plus single mutants, etc.). Color scheme: black dots - training set, orange dots – validation set, blue / rainbow dots – test set containing quintuple mutations (standard used across this work for mutant-based shuffling), cyan dots - test sets containing other mutations.}
  \label{appendix:fig:8}
\end{figure}

\clearpage

\begin{figure}[p]
  \centering
  \includegraphics[width=\textwidth]{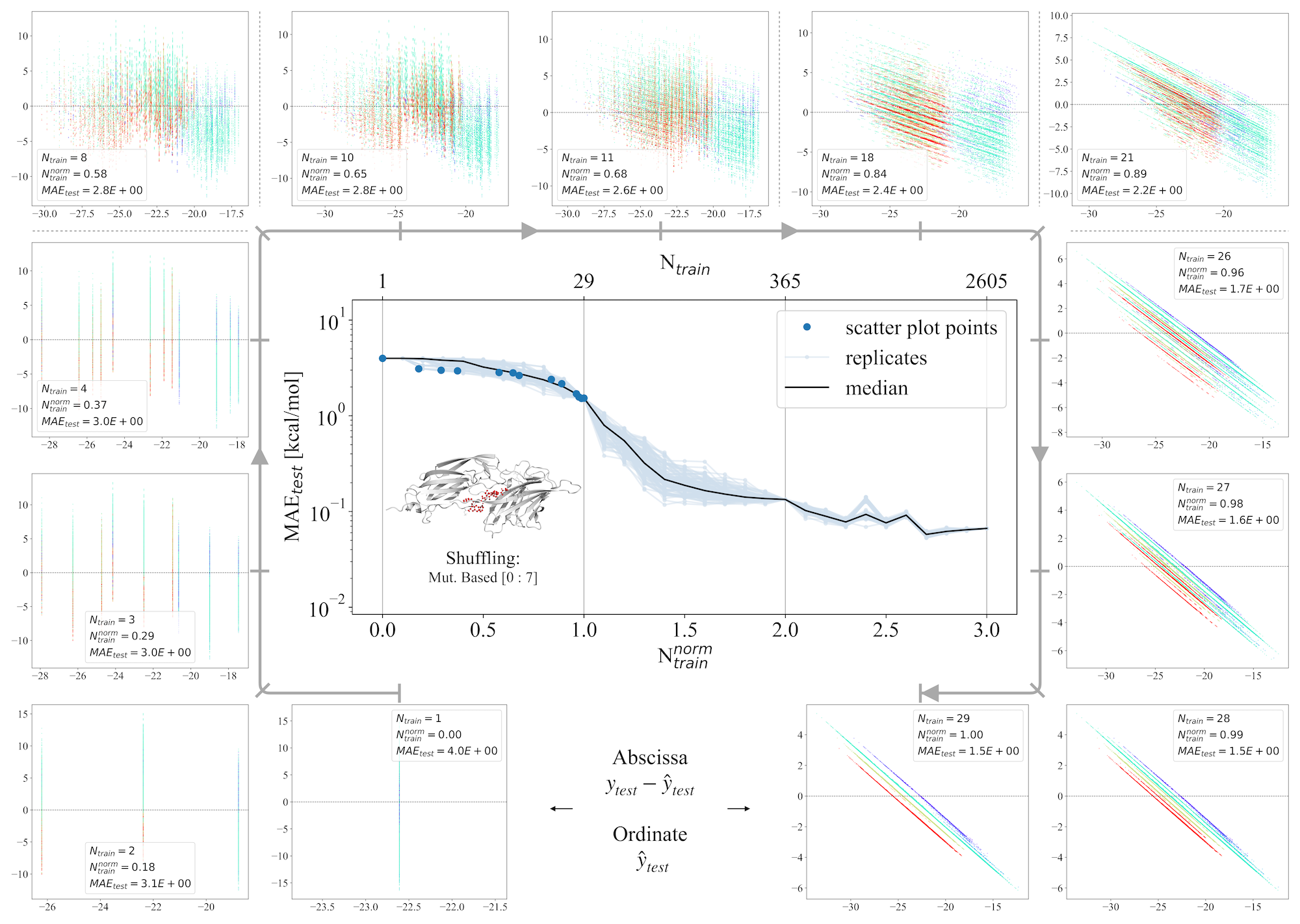}
  \caption{Error vs. predicted extension on data generated with EvoEF function (WT and single mutants) at different training instances. The central plot shows all LCs (100 replicates) and the specific points at which the analysis was carried (same replicate). The colormap is based on the clusters at the point where WT and all single mutants were included in the training set.}
  \label{appendix:fig:9}
\end{figure}

\clearpage

\begin{figure}[p]
  \centering
  \includegraphics[width=\textwidth]{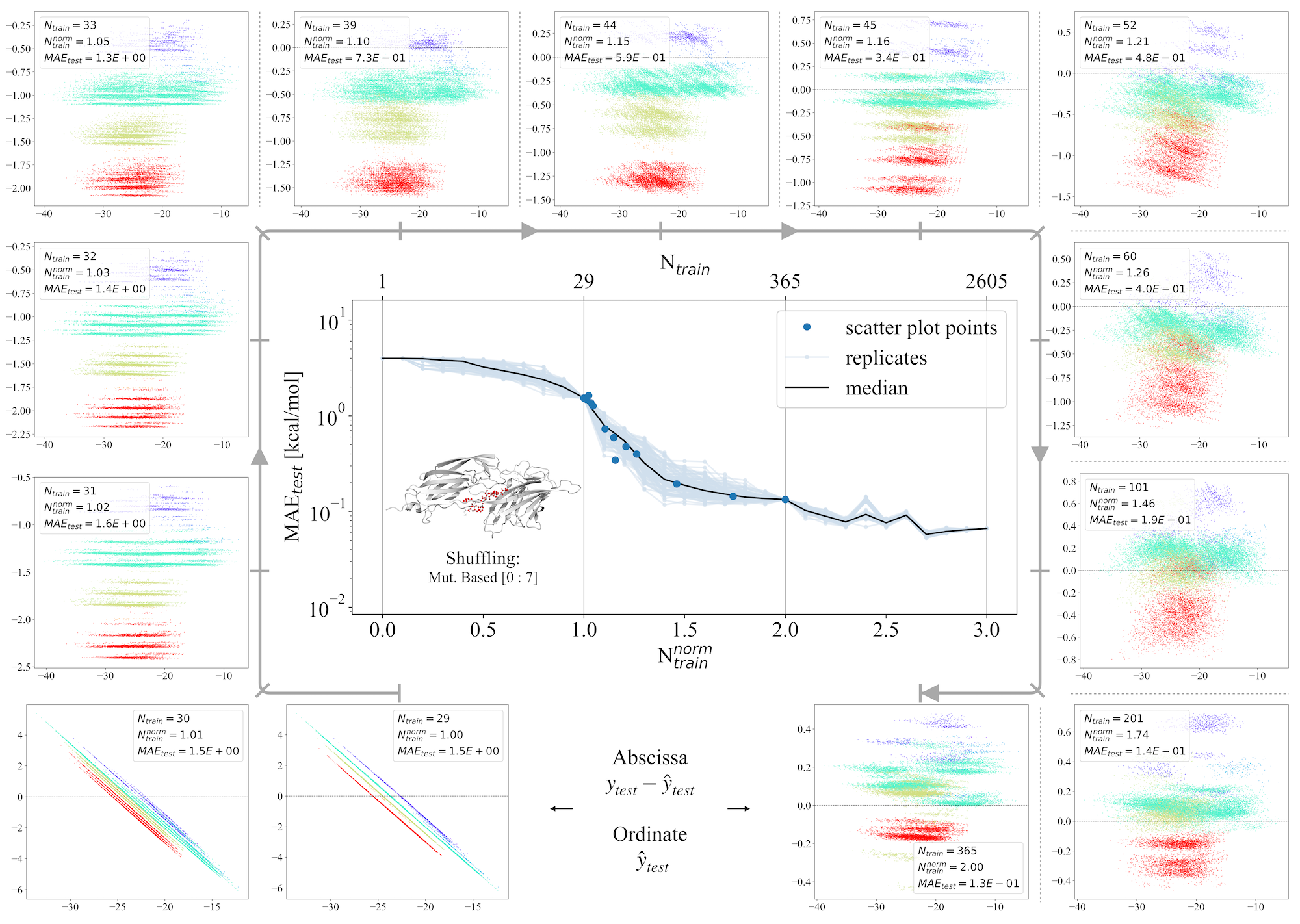}
  \caption{Error vs. predicted extension on data generated with EvoEF function (double mutants) at different training instances. The central plot shows all LCs (100 replicates) and the specific points at which the analysis was carried (same replicate). The colormap is based on the clusters at the point where WT and all single mutants were included in the training set.}
  \label{appendix:fig:10}
\end{figure}

\begin{figure}[p]
  \centering
  \includegraphics[width=\textwidth]{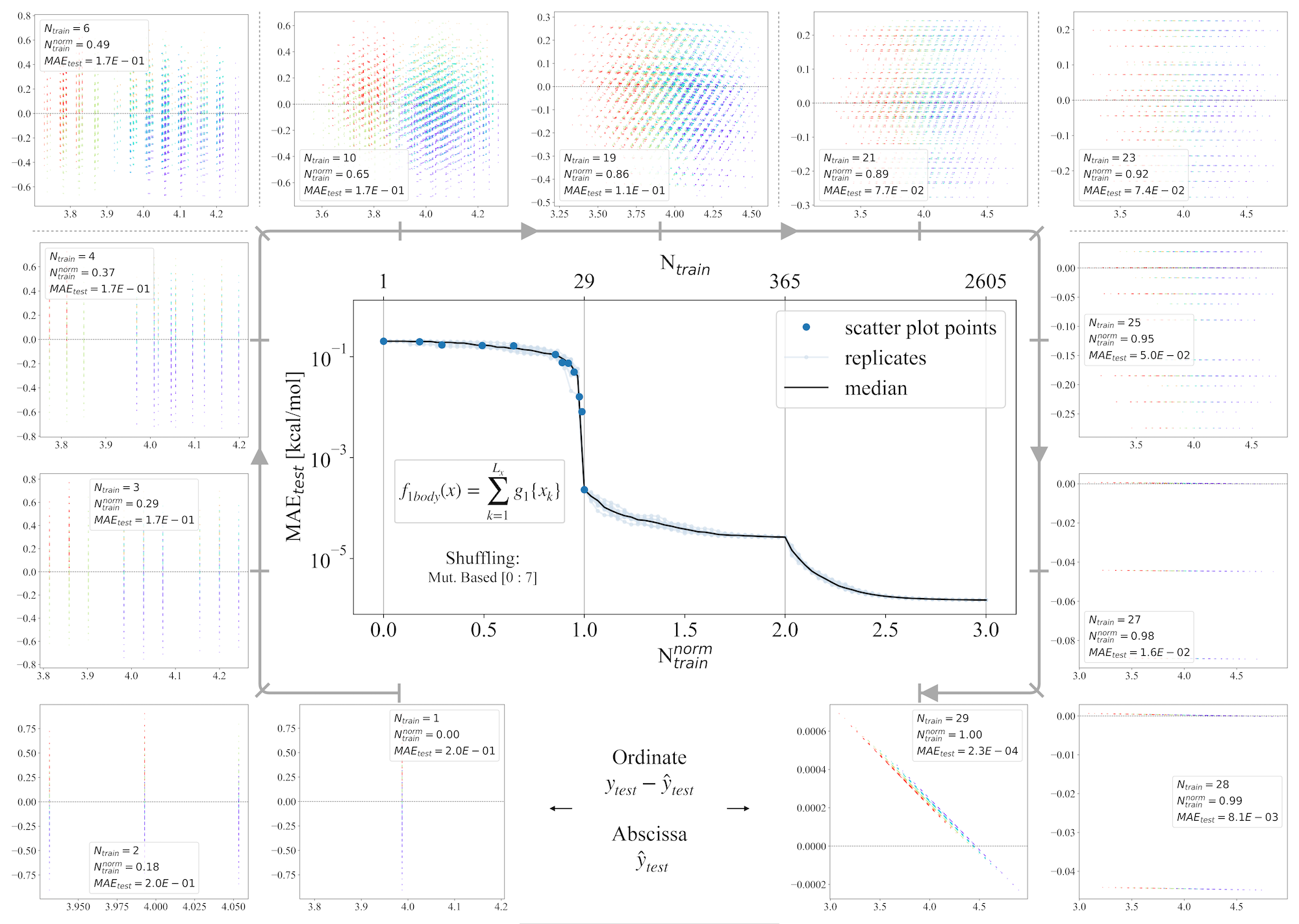}
  \caption{Error vs. predicted extension on data generated with 1-body term function (WT and single mutants) at different training instances. The central plot shows all LCs (100 replicates) and the specific points at which the analysis was carried (same replicate). The colormap is based on the clusters at the point where WT and all single mutants were included in the training set.}
  \label{appendix:fig:11}
\end{figure}

\clearpage

\begin{figure}[p]
  \centering
  \includegraphics[width=\textwidth]{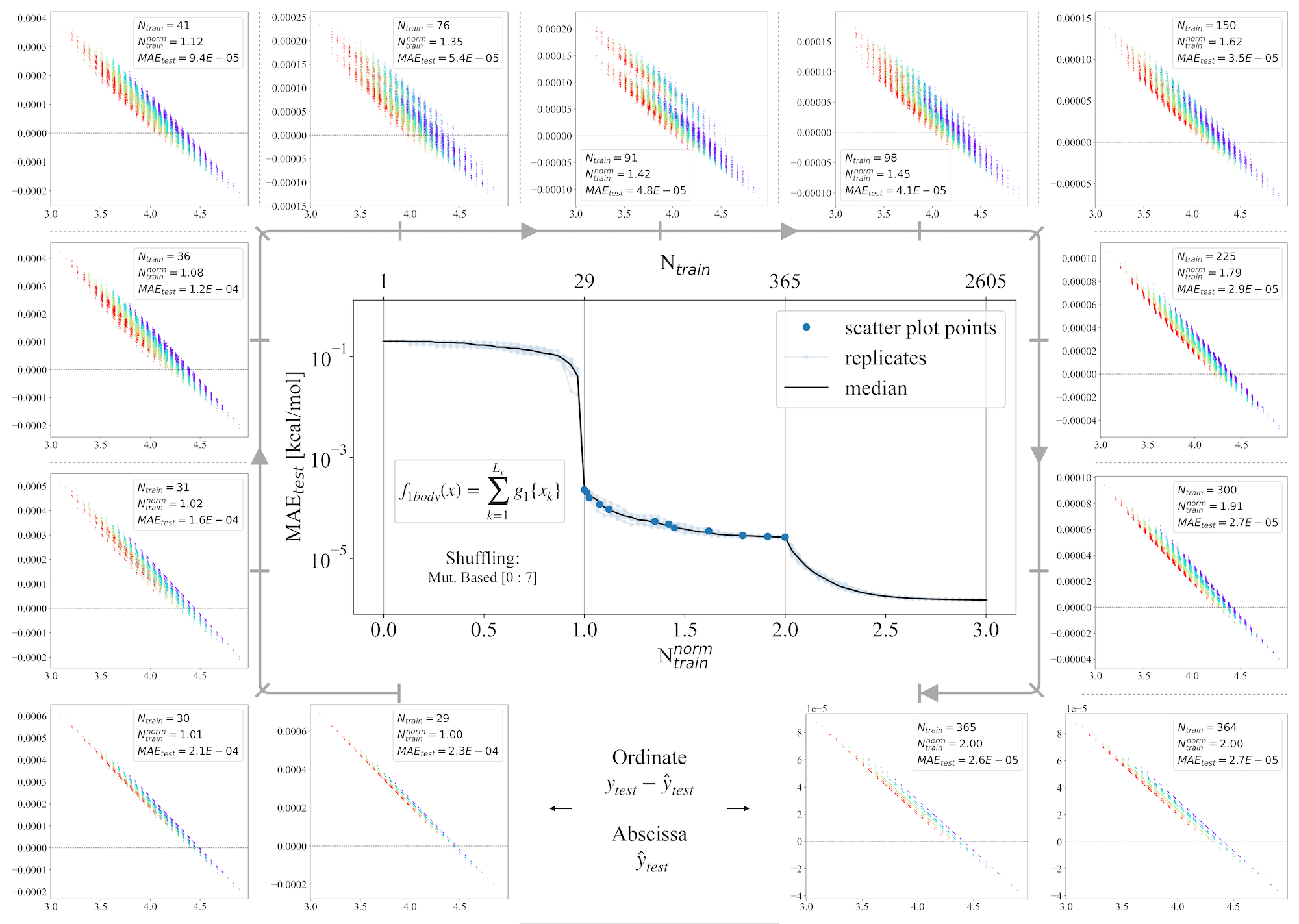}
  \caption{Error vs. predicted extension on data generated with 1-body term function (double mutants) at different training instances. The central plot shows all LCs (100 replicates) and the specific points at which the analysis was carried (same replicate). The colormap is based on the clusters at the point where WT and all single mutants were included in the training set.}
  \label{appendix:fig:12}
\end{figure}

\clearpage

\begin{figure}[p]
  \centering
  \includegraphics[width=\textwidth]{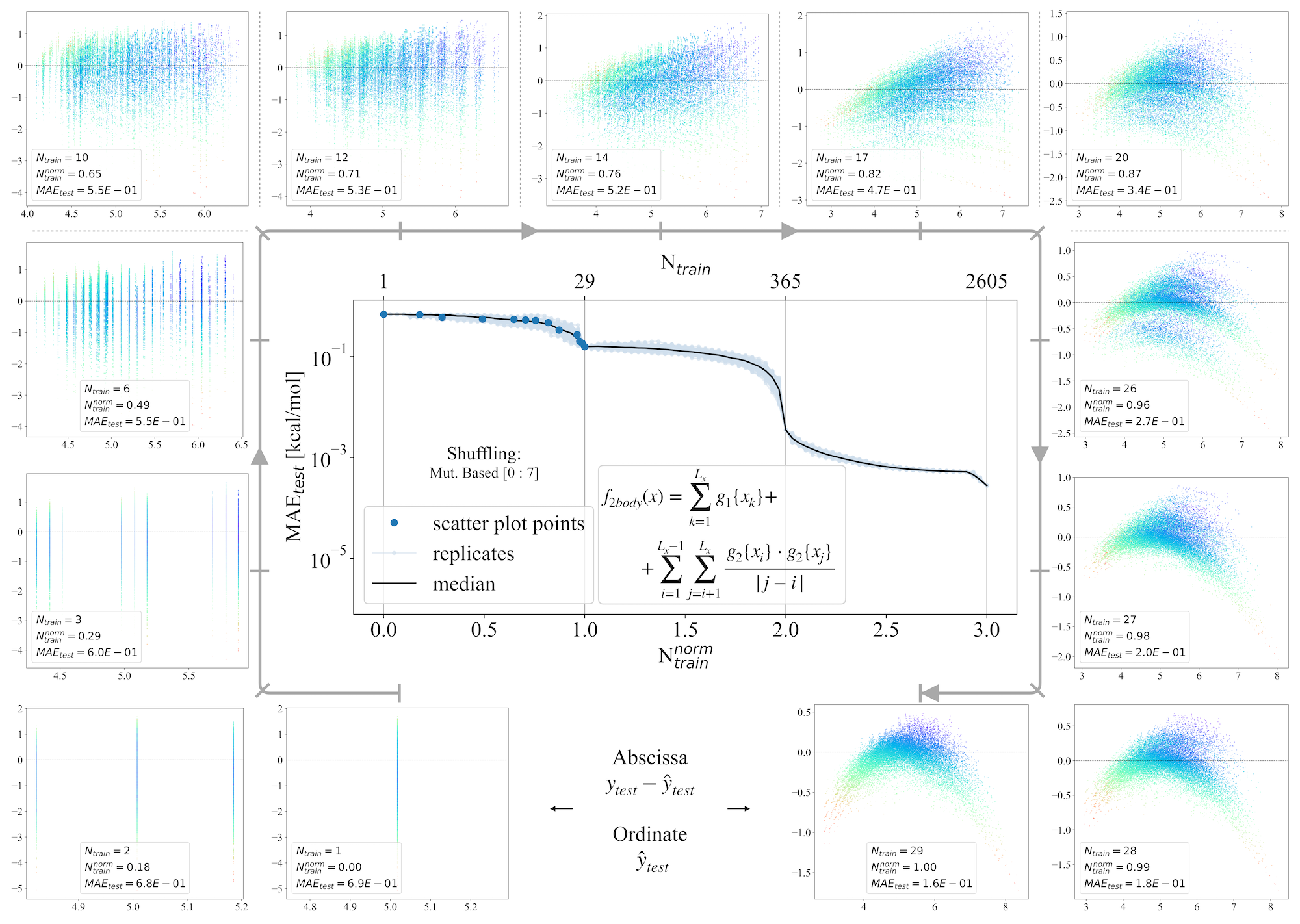}
  \caption{Error vs. predicted extension on data generated with 2-body term function (WT and single mutants) at different training instances. The central plot shows all LCs (100 replicates) and the specific points at which the analysis was carried (same replicate). The colormap is based on the clusters at the point where WT and all single mutants were included in the training set.}
  \label{appendix:fig:13}
\end{figure}

\clearpage

\begin{figure}[p]
  \centering
  \includegraphics[width=\textwidth]{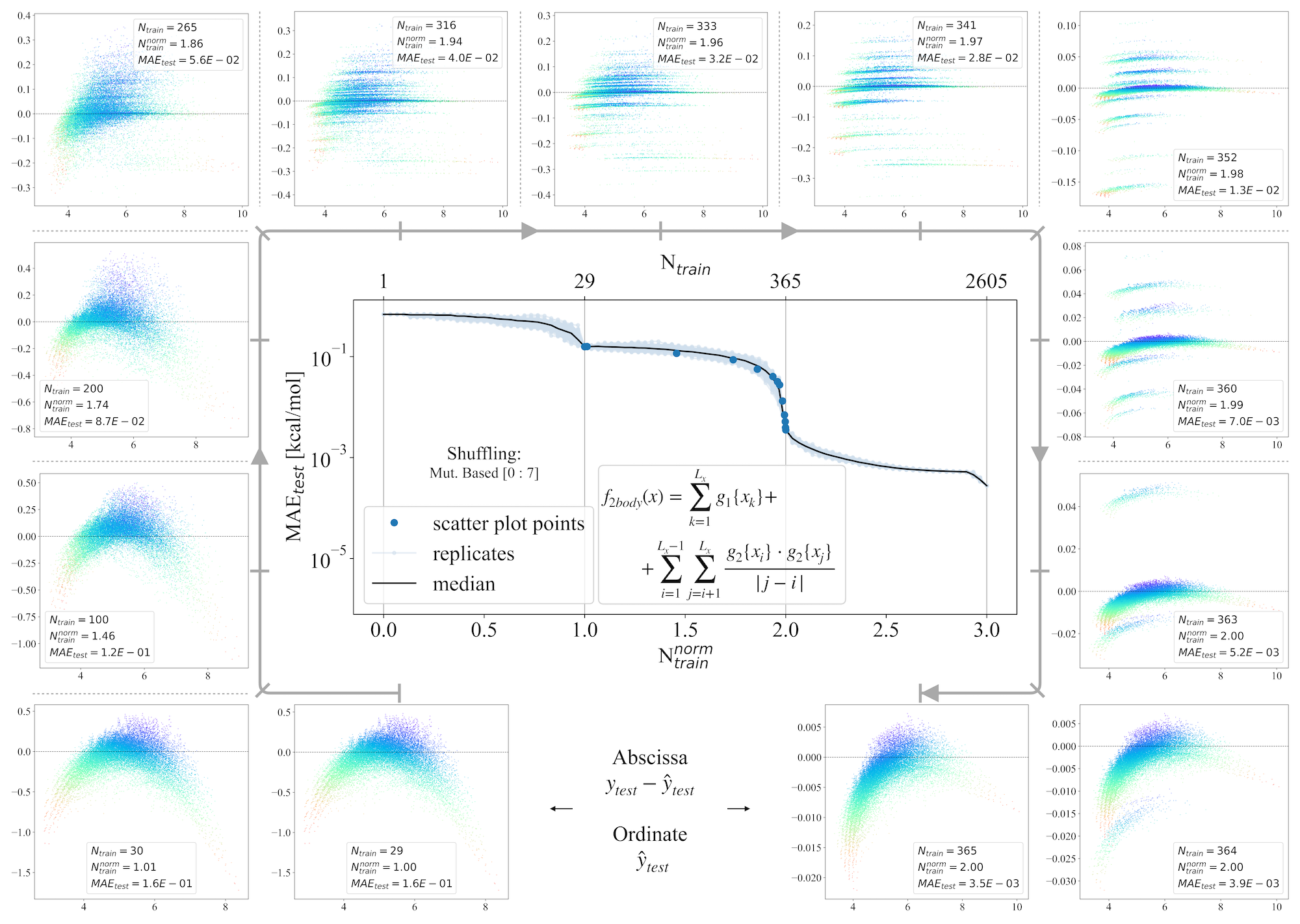}
  \caption{Error vs. predicted extension on data generated with 2-body term function (double mutants) at different training instances. The central plot shows all LCs (100 replicates) and the specific points at which the analysis was carried (same replicate). The colormap is based on the clusters at the point where WT and all single mutants were included in the training set.}
  \label{appendix:fig:14}
\end{figure}

\clearpage

\begin{figure}[p]
  \centering
  \includegraphics[width=\textwidth]{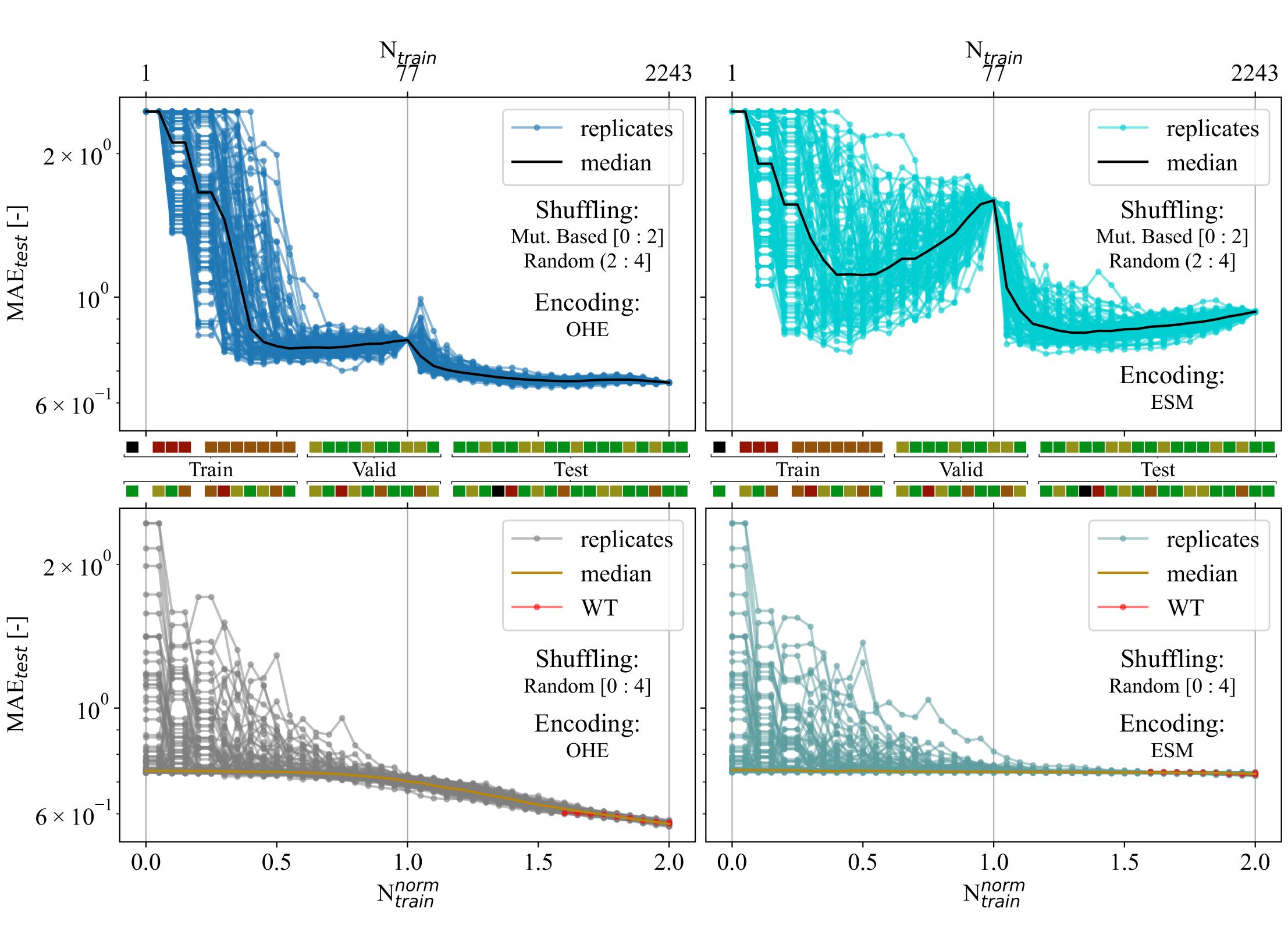}
  \caption{\hl{LCs predicting experimental GB1 fitness with different shuffling (rows) and encoding (columns, from left to right: one hot encoding, and evolutionary scale modeling embeddings).}}
  \label{appendix:fig:15}
\end{figure}

\clearpage

\section{Supporting Tables}

\vspace*{\fill}

\begin{table}[htb!]
  \caption[Size of mutagenised databases generated]{Size of mutagenised databases generated. The top part refers to the specific mutations (WT, single, double, etc.), while the bottom, correspond to the cumulative number of mutations (WT, WT+single, WT+single+double, etc.).}
  \label{tab:DatabaseMutations}
  \centering
\begin{tabular}{lcccccccc}
 \toprule
 \textbf{Database} & $m$=0& $m$=1& $m$=2& $m$=3& $m$=4& $m$=5& $m$=6& $m$=7\\
 \midrule
Fg-$\beta$  & 1 & 28 & 336 & 2240 & 8960 & 21504 & 28672 & 16384 \\
Fg-$\beta$ Ext. & 1 & 49 & 1029 & 12005 & 84035 & 352947 & 823543 & 823543 \\
Chem. Space & 2 & 24 & 120 & 320 & 480 & 384 & 128 & - \\
\hl{GB1 (exp.)} & 1 & 76 & 2091 & 26019 & 121174 & - & - & - \\
\hl{GB1 (imputed)} & 0 & 0 & 75 & 1417 & 9147 & - & - & - \\
\hl{GB1 (tot)} & 1 & 76 & 2166 & 27436 & 130321 & - & - & - \\
\midrule
Fg-$\beta$  & 1 & 29 & 365 & 2605 & 11565 & 33069 & 61741 & 78125 \\
Fg-$\beta$ Ext. & 1 & 50 & 1079 & 13084 & 97119 & 450066 & 1273609 & 2097152 \\
Chem. Space & 2 & 26 & 146 & 466 & 946 & 1330 & 1458 & - \\
\hl{GB1 (tot)} & 1 & 77 & 2243 & 29679 & 160000 & - & - & - \\
  \bottomrule
 \end{tabular}
\end{table}

\vspace*{\fill}

\begin{table}[h]
  \caption{Parameters of dictionaries $g_1\{\cdot\}$ and $g_2\{\cdot\}$.}
  \label{tab:Dictionaries}
  \centering
\begin{tabular}{lll}
 \toprule
 Key & z=1& z=2\\
 \midrule
$g_z$\{A\} & 0.548 & 0.417 \\
$g_z$\{F\} & 0.715 & 0.720 \\
$g_z$\{G\} & 0.602 & 0.000 \\
$g_z$\{R\} & 0.544 & 0.302 \\
$g_z$\{S\} & 0.423 & 0.146 \\
 \midrule
$g_z$\{C\} & 0.645 & 0.092 \\
$g_z$\{D\} & 0.437 & 0.186 \\
$g_z$\{W\} & 0.891 & 0.345 \\
  \bottomrule
 \end{tabular}
\end{table}

\vspace*{\fill}

\clearpage

\begin{table}[p]
  \caption[Z-score binned dictionary]{Z-score binned dictionary.}
  \label{appendix:tab:zscore_binned}
  \centering
  \footnotesize
\begin{tabular}{lccccccccccccccccccc}
 \toprule
 \textbf{AA} & \multicolumn{4}{c}{Lipophilicity} & \multicolumn{4}{c}{Size} & \multicolumn{3}{c}{Polarity}  & \multicolumn{4}{c}{Electronegativity} & \multicolumn{4}{c}{Electrophilicity} \\
 \cmidrule(r){2-5}
\cmidrule(r){6-9}
\cmidrule(r){10-12}
\cmidrule(r){13-16}
\cmidrule(r){17-20}
   & -- & - & + & ++ & -- & - & + & ++ & - & + & ++ & -- & - & + & ++ & -- & - & + & ++ \\
 \midrule
A & 0 & 0 & 1 & 0 & 1 & 0 & 0 & 0 & 0 & 0 & 1 & 0 & 1 & 0 & 0 & 0 & 0 & 0 & 1  \\
C & 0 & 0 & 1 & 0 & 0 & 1 & 0 & 0 & 0 & 0 & 1 & 0 & 0 & 1 & 0 & 1 & 0 & 0 & 0  \\
D & 1 & 0 & 0 & 0 & 0 & 0 & 1 & 0 & 0 & 0 & 1 & 1 & 0 & 0 & 0 & 0 & 0 & 1 & 0  \\
E & 1 & 0 & 0 & 0 & 0 & 0 & 1 & 0 & 0 & 1 & 0 & 1 & 0 & 0 & 0 & 0 & 1 & 0 & 0  \\
F & 0 & 0 & 0 & 1 & 0 & 0 & 0 & 1 & 0 & 0 & 1 & 0 & 0 & 1 & 0 & 0 & 1 & 0 & 0  \\
G & 1 & 0 & 0 & 0 & 1 & 0 & 0 & 0 & 0 & 0 & 1 & 0 & 1 & 0 & 0 & 0 & 1 & 0 & 0  \\
H & 1 & 0 & 0 & 0 & 0 & 0 & 0 & 1 & 0 & 0 & 1 & 0 & 0 & 0 & 1 & 0 & 0 & 1 & 0  \\
I & 0 & 0 & 0 & 1 & 0 & 1 & 0 & 0 & 1 & 0 & 0 & 0 & 1 & 0 & 0 & 0 & 0 & 1 & 0  \\
K & 1 & 0 & 0 & 0 & 0 & 0 & 1 & 0 & 1 & 0 & 0 & 0 & 0 & 1 & 0 & 0 & 0 & 1 & 0  \\
L & 0 & 0 & 0 & 1 & 0 & 1 & 0 & 0 & 0 & 1 & 0 & 0 & 1 & 0 & 0 & 0 & 0 & 1 & 0  \\
M & 0 & 0 & 0 & 1 & 0 & 1 & 0 & 0 & 0 & 0 & 1 & 0 & 0 & 0 & 1 & 0 & 1 & 0 & 0  \\
N & 1 & 0 & 0 & 0 & 0 & 0 & 1 & 0 & 0 & 0 & 1 & 0 & 1 & 0 & 0 & 0 & 0 & 0 & 1  \\
P & 0 & 1 & 0 & 0 & 0 & 0 & 1 & 0 & 0 & 0 & 1 & 0 & 0 & 1 & 0 & 0 & 0 & 0 & 1  \\
Q & 0 & 0 & 1 & 0 & 0 & 0 & 1 & 0 & 0 & 1 & 0 & 1 & 0 & 0 & 0 & 0 & 0 & 1 & 0  \\
R & 1 & 0 & 0 & 0 & 0 & 0 & 0 & 1 & 1 & 0 & 0 & 0 & 0 & 0 & 1 & 0 & 1 & 0 & 0  \\
S & 1 & 0 & 0 & 0 & 0 & 1 & 0 & 0 & 0 & 0 & 1 & 1 & 0 & 0 & 0 & 0 & 0 & 1 & 0  \\
T & 0 & 0 & 1 & 0 & 1 & 0 & 0 & 0 & 0 & 1 & 0 & 1 & 0 & 0 & 0 & 0 & 1 & 0 & 0  \\
V & 0 & 0 & 0 & 1 & 1 & 0 & 0 & 0 & 0 & 1 & 0 & 0 & 1 & 0 & 0 & 0 & 1 & 0 & 0  \\
W & 0 & 0 & 0 & 1 & 0 & 0 & 0 & 1 & 0 & 0 & 1 & 0 & 0 & 0 & 1 & 1 & 0 & 0 & 0  \\
Y & 0 & 0 & 0 & 1 & 0 & 0 & 0 & 1 & 0 & 0 & 1 & 0 & 1 & 0 & 0 & 1 & 0 & 0 & 0  \\
  \bottomrule
 \end{tabular}
\end{table}

\afterpage{%
    \clearpage
    \thispagestyle{empty}
    \begin{landscape}

\begin{table}
  \caption[AAindex PCA reduced (18 principal components) dictionary]{AAindex PCA reduced (18 principal components) dictionary. Each column was normalized by a min-max scaler.}
  \label{appendix:tab:aaindex}
  \centering
  \footnotesize
\begin{tabular}{ccccccccccccccccccc}
 \toprule
 \textbf{AA} & PC0 & PC1 & PC2 & PC3 & PC4 & PC5 & PC6 & PC7 & PC8 & PC9 & PC10 & PC11 & PC12 & PC13 & PC14 & PC15 & PC16 & PC17 \\
   & 34.4\% & 15.8\% & 11.7\% & 7.0\% & 5.5\% & 4.5\% & 3.2\% & 2.5\% & 2.2\% & 2.1\% & 1.9\% & 1.8\% & 1.5\% & 1.2\% & 1.1\% & 1.1\% & 1.0\% & 0.7\% \\
 \midrule
A & 0.476 & 0.574 & 1.000 & 0.479 & 0.629 & 0.590 & 0.210 & 0.352 & 0.624 & 0.242 & 0.511 & 0.568 & 0.520 & 0.626 & 0.981 & 0.853 & 0.645 & 0.181  \\
R & 0.261 & 0.000 & 0.432 & 0.453 & 0.000 & 0.012 & 0.237 & 0.036 & 0.141 & 0.000 & 0.863 & 0.308 & 0.200 & 0.464 & 0.647 & 0.419 & 0.418 & 0.415  \\
N & 0.080 & 0.421 & 0.335 & 0.272 & 0.354 & 0.658 & 0.601 & 0.753 & 0.317 & 0.720 & 0.605 & 0.313 & 0.006 & 0.104 & 1.000 & 0.484 & 0.321 & 0.231  \\
D & 0.000 & 0.344 & 0.466 & 0.381 & 0.761 & 0.788 & 1.000 & 0.304 & 0.156 & 0.303 & 0.545 & 0.357 & 0.123 & 1.000 & 0.185 & 0.471 & 0.695 & 0.403  \\
C & 0.681 & 0.642 & 0.238 & 0.000 & 1.000 & 0.000 & 0.325 & 0.070 & 0.218 & 0.678 & 0.635 & 0.494 & 0.477 & 0.419 & 0.401 & 0.487 & 0.424 & 0.351  \\
Q & 0.269 & 0.179 & 0.526 & 0.430 & 0.531 & 0.339 & 0.367 & 0.424 & 0.366 & 0.420 & 0.000 & 0.212 & 0.542 & 0.000 & 0.407 & 0.302 & 1.000 & 0.466  \\
E & 0.161 & 0.110 & 0.829 & 0.522 & 0.886 & 0.678 & 0.737 & 0.180 & 0.104 & 0.321 & 0.442 & 0.616 & 0.644 & 0.076 & 0.531 & 0.351 & 0.000 & 0.486  \\
G & 0.050 & 1.000 & 0.512 & 0.285 & 0.034 & 1.000 & 0.076 & 0.119 & 0.019 & 0.269 & 0.343 & 0.482 & 0.411 & 0.345 & 0.289 & 0.316 & 0.385 & 0.415  \\
H & 0.462 & 0.199 & 0.310 & 0.305 & 0.504 & 0.520 & 0.172 & 1.000 & 0.067 & 0.193 & 0.585 & 1.000 & 0.461 & 0.546 & 0.417 & 0.291 & 0.547 & 0.439  \\
I & 1.000 & 0.579 & 0.556 & 0.552 & 0.312 & 0.366 & 0.688 & 0.395 & 0.000 & 0.481 & 0.182 & 0.550 & 0.000 & 0.389 & 0.645 & 0.374 & 0.426 & 0.392  \\
L & 0.929 & 0.510 & 0.836 & 0.634 & 0.382 & 0.587 & 0.413 & 0.452 & 0.464 & 0.722 & 1.000 & 0.332 & 0.483 & 0.488 & 0.296 & 0.000 & 0.537 & 0.327  \\
K & 0.171 & 0.063 & 0.653 & 0.491 & 0.139 & 0.365 & 0.152 & 0.321 & 0.414 & 1.000 & 0.192 & 0.618 & 0.298 & 0.718 & 0.144 & 0.656 & 0.276 & 0.412  \\
M & 0.885 & 0.295 & 0.569 & 0.372 & 0.815 & 0.590 & 0.000 & 0.619 & 0.190 & 0.198 & 0.142 & 0.000 & 0.245 & 0.758 & 0.512 & 0.336 & 0.124 & 0.422  \\
F & 0.968 & 0.410 & 0.339 & 0.515 & 0.394 & 0.713 & 0.476 & 0.551 & 0.225 & 0.306 & 0.710 & 0.312 & 0.379 & 0.091 & 0.000 & 1.000 & 0.423 & 0.562  \\
P & 0.039 & 0.722 & 0.000 & 1.000 & 0.841 & 0.266 & 0.204 & 0.399 & 0.261 & 0.438 & 0.472 & 0.476 & 0.350 & 0.476 & 0.537 & 0.453 & 0.400 & 0.399  \\
S & 0.146 & 0.638 & 0.511 & 0.339 & 0.339 & 0.317 & 0.550 & 0.616 & 0.850 & 0.403 & 0.532 & 0.329 & 0.506 & 0.594 & 0.641 & 0.421 & 0.317 & 1.000  \\
T & 0.344 & 0.583 & 0.473 & 0.398 & 0.316 & 0.127 & 0.695 & 0.571 & 0.923 & 0.080 & 0.291 & 0.474 & 0.376 & 0.362 & 0.042 & 0.375 & 0.207 & 0.000  \\
W & 0.921 & 0.187 & 0.045 & 0.437 & 0.502 & 0.965 & 0.383 & 0.000 & 1.000 & 0.344 & 0.356 & 0.678 & 0.196 & 0.420 & 0.629 & 0.321 & 0.446 & 0.482  \\
Y & 0.683 & 0.334 & 0.050 & 0.451 & 0.070 & 0.590 & 0.739 & 0.407 & 0.103 & 0.492 & 0.338 & 0.301 & 1.000 & 0.737 & 0.755 & 0.484 & 0.366 & 0.242  \\
V & 0.879 & 0.655 & 0.664 & 0.485 & 0.292 & 0.118 & 0.730 & 0.325 & 0.106 & 0.310 & 0.224 & 0.684 & 0.298 & 0.518 & 0.605 & 0.459 & 0.453 & 0.616  \\
  \bottomrule
 \end{tabular}
\end{table}
    \end{landscape}
    \clearpage
}

\clearpage

\putbib 
\end{bibunit}

\end{document}